\documentclass[aoas]{imsart}

\RequirePackage{amsthm,amsmath,amsfonts,amssymb}
\RequirePackage[authoryear]{natbib}
\RequirePackage[colorlinks,citecolor=blue,urlcolor=blue]{hyperref}
\RequirePackage{graphicx}
\RequirePackage{multirow}
\RequirePackage{xcolor}
\RequirePackage{float}

\RequirePackage{dsfont}
\RequirePackage{listings}

\startlocaldefs
\theoremstyle{plain}
\newtheorem{theorem}{Theorem}[section]
\newtheorem{proposition}[theorem]{Proposition}

\DeclareMathOperator*{\argmin}{arg\,min}

\endlocaldefs

\begin{document}

\begin{frontmatter}
\title{Quantifying sleep apnea heterogeneity using hierarchical Bayesian modeling}
\runtitle{Quantifying sleep apnea heterogeneity}

\begin{aug}
\author[A]{\fnms{Glenn}~\snm{Palmer}\ead[label=e1]{glennpalmer55@gmail.com}\orcid{0009-0009-7138-4914}},
\author[B]{\fnms{Narat}~\snm{Srivali}\ead[label=e2]{narat.srivali@duke.edu}\orcid{0000-0002-6945-329X}}
\and
\author[A]{\fnms{David}~\snm{B. Dunson}\ead[label=e3]{dunson@duke.edu}\orcid{0000-0003-4942-1597}}
\address[A]{Department of Statistical Science,
Duke University\printead[presep={,\ }]{e1,e3}}
\address[B]{Division of Pulmonary, Allergy, and Critical Care Medicine, Duke University School of Medicine\printead[presep={,\ }]{e2}}
\end{aug}

\begin{abstract}

Obstructive Sleep Apnea (OSA) is a breathing disorder during sleep that affects millions of people worldwide. The diagnosis of OSA often occurs through an overnight polysomnogram (PSG) sleep study that generates a massive amount of physiological data. However, despite the evidence of substantial heterogeneity in the expression and symptoms of OSA, diagnosis and scientific analysis of severity typically focus on a single summary statistic, the Apnea-Hypopnea Index (AHI). We address the limitations of this approach through hierarchical Bayesian modeling of PSG data. Our approach produces interpretable random effects for each patient, which govern sleep-stage dynamics, rates of OSA events, and  impacts of OSA events on subsequent sleep-stage dynamics. We propose a novel approach for using these random effects to produce a Bayes optimal clustering of patients. We use the proposed approach to analyze data from the APPLES study. Our analysis produces clinically interesting groups of patients with sleep apnea and a novel finding of an association between OSA expression and cognitive performance that is missed by an AHI-based analysis.

\end{abstract}


\end{frontmatter}

\section{Introduction}\label{section_intro}

Obstructive sleep apnea (OSA) is a widespread breathing disorder characterized by periodic narrowing or obstruction of the pharyngeal airway of a person during sleep, leading to intervals of decreased or stopped breathing \citep{osman2018obstructive}. Estimates of OSA prevalence in the general population range from $9\%$ to $38\%$, and are higher for particular subgroups such as men, older adults, and people with obesity \citep{senaratna2017prevalence}. Associations have been found between untreated OSA and health conditions, including metabolic disorders \citep{drager2013obstructive}, cardiovascular disease \citep{kapur2008sleep, walia2014association}, and depression \citep{wheaton2012sleep}. However, despite the potentially severe consequences of OSA and its high prevalence, most people with OSA are not diagnosed \citep{osman2018obstructive}, suggesting that improving the OSA diagnosis process is an important public health issue.

The gold standard for diagnosing OSA is the polysomnogram (PSG), in which a patient spends a night in a sleep lab, and their airflow, respiratory effort, pulse, oxygen saturation, movements, and sleep stages (as indicated by EEG analysis) are recorded \citep{rundo2019polysomnography}. These PSG data are analyzed to identify events when a person's breathing decreased significantly or stopped for at least 10 seconds, referred to as hypopnea or apnea events, depending on their severity. The total count of these events is divided by the total number of hours of sleep to calculate the apnea-hypopnea index (AHI), which is used to diagnose OSA -- an AHI above 5 indicates sleep apnea, with higher values interpreted as more severe cases \citep{rundo2019polysomnography}. Although the AHI is simple to compute and interpret, it has serious limitations. \cite{osman2018obstructive} note that AHI misses important differences in event severity; a person with low AHI can experience significant effects in terms of oxygen desaturation and clinical symptoms if their apnea events are severe. On the other hand, if a person has a large number of minor events that are not extreme enough to be scored, they could have a misleadingly low AHI compared to the extent to which their breathing truly is disordered \citep{sankari2017characteristics, koch2017breathing}. There are other sources of heterogeneity within patients who have a given AHI, such as which stages of sleep the events occur during and how disruptive they were to sleep continuity. Given these factors, studies have often struggled to find meaningful associations between AHI and clinical symptoms \citep{weaver2005polysomnography}. Given the time and expense involved in performing PSG studies and the rich datasets they produce, it is critical to obtain more robust, descriptive, and clinically useful measures of OSA.

In this article, we propose an approach to illuminate the relationship between the rate of events of apnea and hypopnea and sleep stage dynamics in people with OSA. There has been recent interest in sleep stage-specific OSA, with some evidence that a high rate of events during rapid eye movement (REM) sleep may be particularly harmful \citep{aurora2018obstructive, varga2019rem}. There is also a great interest in the fraction of time people spend in each stage of sleep, often referred to as sleep architecture \citep{pase2023sleep}. However, there have been limited approaches to unravel the relationship between the two \citep{bianchi2010obstructive}. We seek to develop a modeling approach that can (1) illuminate population-level variation in sleep stage dynamics and sleep stage-specific OSA, (2) identify individuals whose sleep dynamics are particularly strongly affected by their apnea and hypopnea events, and (3) evaluate whether the sleep-stage disruptiveness of a case of OSA is associated with other clinical variables, including the severity of symptoms. To do so, we formulate a hierarchical Bayesian model, with patient-level random effects governing the variation in stage-specific event rates and stage-transition probabilities.

\subsection{Relevant literature}\label{relevant_literature}

It has been widely acknowledged in the literature on sleep research that there is substantial heterogeneity in patients with OSA that is not represented by a simple AHI-based approach. \cite{eckert2013defining} proposed a framework for defining OSA ``phenotypes'' based on anatomical characteristics and the impact of OSA on breathing patterns and arousal, suggesting that different treatments may be optimal for different phenotypes. \cite{mokhlesi2014obstructive} examined associations between AHI values specific for sleep stages and hypertension, finding that REM AHI greater than 15 was associated with the risk of hypertension, even in patients with a total AHI below 5. \cite{gagnadoux2016relationship} performed a cluster analysis based on patient demographics and symptoms and found that the effectiveness of CPAP treatment varied between clusters. See \cite{zinchuk2017phenotypes} for a clinical review of approaches for OSA phenotyping.

When investigating sleep architecture, simple summary statistics are typically used, such as the total time spent asleep, the amount of time awake after the first onset of sleep, and the amount or fraction of time spent in each sleep stage \citep{hermans2022representations}. However, since such summaries may miss significant variation in dynamics (e.g., a large fraction of REM sleep could result from many short cycles or a small number of long cycles), a smaller literature has modeled sleep stage dynamics directly by (1)
survival analysis-style methods for sleep stage durations or (2) Markov models for transitions between sleep stages
\citep{hermans2022representations}. In the first category, \cite{norman2006sleep} compared ``sleep continuity,'' defined as the duration of intervals of sleep between waking up, between healthy participants and those with OSA using Kaplan-Meier curves \citep{kaplan1958nonparametric} and a log-rank test \citep{mantel1966evaluation}, concluding that the OSA patients had significantly less continuous sleep. \cite{klerman2013survival} modeled REM and non-REM uninterrupted durations with an exponential distribution and found that non-REM stability was higher for younger participants. Also modeling REM and non-REM durations with exponential distributions, \cite{bianchi2010obstructive} found that expected continuous durations were shorter in both REM and non-REM sleep for OSA compared to healthy participants. In the second category, \cite{karlsson2000pharmacodynamic} treated sleep-stage transitions as Markovian and modeled each possible transition using a mixed-effects logistic regression model, applying this approach to evaluate the effectiveness of temazepam for patients with insomnia. \cite{bizzotto2010multinomial} expanded on this approach, using multinomial logistic regressions to ensure the estimated transition probabilities from a given stage added up to 1, and \cite{kjellsson2011modeling} applied this approach to model sleep dynamics in insomnia patients. Finally, \cite{yetton2018quantifying} combined the two types of approaches, modeling sleep stage durations separately from Markovian transition probabilities that excluded self-transitions, using a Bayesian network approach to evaluate how durations and transition probabilities varied by categorical covariates.

Other application areas have motivated models that have similarities with our approach.
\cite{cho2021markov} developed a Markov mixed-effect multinomial logistic regression for dynamically evolving human linguistic sentence structure. \cite{sarkar2018bayesian} developed a hierarchical Markov model for mouse vocalization patterns that depend on genotype and social context.
\cite{zhang2014bayesian} developed a Bayesian mixed hidden Markov model, building on the hierarchical hidden Markov model of \cite{altman2007mixed}, and applied it to modeling dynamic transitions through latent health states jointly with patient questionnaire data. There is also relevant work on hierarchical modeling of multivariate longitudinal data. \cite{hauser2024bayesian} developed a Bayesian longitudinal generalized linear mixed model for high-dimensional clinical data in cancer patients. \cite{damone2025bayesian} developed a joint model for recurrent events, terminal events, and longitudinal outcome data, with dependence among the three induced by a multivariate normal distribution for patient-level random effects.

\subsection{Our contribution}

The above OSA literature conveys a broad recognition of substantial heterogeneity among OSA patients, including their sleep dynamics, which is not captured by AHI, while the parallel literature on joint modeling of longitudinal multivariate clinical data illustrates the utility of such approaches for understanding the relationships among complex data. However, to our knowledge, no one has presented an approach to model the variability of sleep dynamics jointly with the occurrence of OSA events. In particular, existing modeling approaches for analyzing PSG data do not allow the probability of a sleep transition to depend on the occurrence of a recent apnea event. We propose an approach that not only achieves these goals, but also quantifies patient-level heterogeneity in sleep-stage dynamics, in sleep-stage specific rates of OSA events, and critically, in the level of disruptiveness of a patients' OSA events to their sleep-stage dynamics. We then propose an approach for inferring groups of patients with similar random effects that is Bayes optimal under K-means loss. This is a novel, broadly applicable, and easy-to-implement methodological result that will likely be of independent interest to many readers. Our full approach illuminates previously unexplored variation in OSA severity. In doing so, we establish a new framework for modeling OSA heterogeneity that can be applied to numerous cohorts of patients.

In Section \ref{section_data}, we give more information on the motivating application and the data set. In Section \ref{section_model}, we describe our modeling approach and subsequent clustering method. In Section \ref{section_simulations}, we demonstrate the efficacy of our full approach in simulations. In Section \ref{section_application}, we present our data analysis results. Section \ref{section_discussion} presents some discussion and future directions.

\section{Polysomnogram data analysis}\label{section_data}

\subsection{Background}

The Apnea Positive Pressure Long-term Efficacy Study (APPLES) was a multicenter study of the efficacy of continuous positive airway pressure (CPAP) in improving neurocognitive function in OSA patients \citep{kushida2006apnea}. APPLES enrolled 1204 participants, which they randomized into CPAP treatment and control groups and followed for six months. Baseline PSG data measured at each participant's diagnostic visit is available through the National Sleep Research Resource \citep{zhang2018national} for 1104 of these participants, as well as demographic covariates, results of neurocognitive testing, and summaries of follow-up appointments \citep{quan2011association}.

The primary outcomes with measurements at the diagnostic visit in APPLES were two neurocognitive tests -- the Pathfinder Number Test \citep{partington1949partington} and the Buschke Verbal Selective Reminding Test (BSRT) \citep{buschke1973selective}. In Pathfinder, participants try to select consecutive numbers displayed in boxes in different quadrants of a computer screen. Pathfinder is designed to assess attention and psychomotor function, and the primary measured outcome is the time a participant takes to complete the test, up to 60 seconds \citep{quan2011association}. In the BSRT, participants are presented with a list of 12 unrelated words and then asked to recall as many of them as possible over the course of 6 trials. The BSRT is designed to assess verbal learning and short- and long-term memory, and the primary measured outcome is the total number of words correctly recalled over the 6 trials, ranging up to all 72 words \citep{quan2011association}. In their analysis of these test results, \cite{quan2011association} performed hypothesis tests comparing Pathfinder and BSRT scores for participants with baseline AHI values below 10 to those of patients with AHI values above 50. They found that patients with high-AHI performed worse. They then fitted subsequent regression models that also accounted for demographic characteristics, and the associations between the tests and AHI were no longer significant. These results correspond to the general state of the literature, in that there are examples of studies that find associations between OSA and neurocognitive outcomes (e.g., \cite{jokic1999positional} found that Pathfinder scores improved in patients after OSA treatment), but in general the results have been mixed, with other studies not finding associations with cognitive outcomes \citep{weaver2005polysomnography}.

In this work, we postulate that associations between the rate of OSA events and neurocognitive outcomes may vary by how disruptive a particular patient's events are to their sleep dynamics. This corresponds to the increasing focus on personalized approaches to medicine \citep{mathur2017personalized}, which has been suggested to be critical for understanding and quantifying OSA \citep{edwards2019more}. To investigate this possible patient heterogeneity, we propose a hierarchical Bayesian approach to quantify patient-level OSA disruptiveness, and apply it to re-analyze the APPLES baseline PSG data. 

\subsection{Data description and preprocessing}

The released APPLES data set included PSG data for 1104 participants, selected based on being at least 18 years old and having a clinical diagnosis of OSA, but never having used a CPAP before. Other exclusion criteria can be found in \cite{kushida2006apnea}, and include a number of medical conditions, medications, and a history of a sleepiness-related motor vehicle accident \citep{quan2011association}. After enrollment, the participants attended a diagnostic visit, where the PSG was measured between 7 and 9 hours, followed by a day of neurocognitive testing and monitoring. For PSG, sleep stages were manually recorded for 30-second intervals referred to as epochs using the Rechtschaffen and Kales criteria \citep{rechtschaffen1968manual}, categorized into awake, REM sleep, and three levels of non-REM sleep numbered by depth. Apnea events were defined by a decrease of at least 90\% from the baseline nasal pressure amplitude for an interval of at least 10 seconds. Hypopnea events were defined as a 10-second or longer decrease in nasal pressure between 50\% and 90\%, or a decrease less than 50\% that was also accompanied by a 3\% or greater oxygen desaturation or arousal, according to \citep{american1999sleep}. The apnea hypopnea index (AHI) was calculated as the total number of apnea and hypopnea events divided by the total hours of sleep.

Before analysis, we removed nine patients who never entered REM sleep, as they likely had unusually poor sleep during the testing procedure. We removed an additional two patients whose recorded sleep stages had inconsistencies in the files for their PSG data, leaving 1093 patients for our primary analysis. In our follow-up analyses of the Pathfinder and BSRT tests, two and one participants were missing these scores, respectively, and therefore were excluded from the corresponding second-stage analysis only.

\section{Model}\label{section_model}

\subsection{Model description}\label{Model_description}

Consider $n$ patients, each monitored for a single night of sleep. We model each patient's epoch-to-epoch sleep stage transitions as probabilities of going from stage $s_o \in \{\text{REM, non-REM}\}$ in one epoch to stage $s_n \in \{\text{Awake, REM, non-REM}\}$ in the next. The subscripts in $s_o$ and $s_n$ represent the ``old'' and ``new'' stages, respectively. We allow the probability of each transition to depend on whether an apnea or hypopnea event is occurring in the current epoch, and we allow both the baseline (no-event) transition probabilities and the effects of having an event to vary among patients. By restricting $s_o$ to REM or non-REM, we include transitions \textit{to} being awake, but exclude transitions \textit{from} being awake in our model. This is clinically appropriate because when reviewing PSGs, awakening patterns reveal OSA severity, while how patients return to sleep provides little information about OSA pathophysiology; it is instead driven by sleep cycle phase, behavioral factors (bathroom use, anxiety), environmental factors, and comorbidities.

To formalize the above, let $s_{ij} = s_o \in \{\text{REM, non-REM}\}$ be the sleep stage for the $i$th patient during the $j$th epoch, where $j \in \{1,...,m_i\}$ given that patient $i$ is observed for $m_i$ epochs. We model transition probabilities to $s_{i,j+1} = s_n \in \{\text{Awake, REM, non-REM}\}$ using a mixed effects multinomial logistic regression model,
$$Pr(s_{i,j+1}=s_n | s_{ij}=s_o) = 
\frac{\exp(\delta_{ijs_os_n})}{\sum_{s'_n} \exp(\delta_{ijs_os'_n})}$$
with
$$\delta_{ijs_os_n} = 
\begin{cases}
    \mu_{s_os_n} + \gamma_{i s_os_n} + (\tau_{s_os_n} + \alpha_{i s_os_n}) \cdot v_{ij} & \text{ if } s_o \neq s_n \\
    0 & \text{ if } s_o = s_n
\end{cases}$$
where $v_{ij} = \mathds{1} (\text{patient $i$ has event during epoch $j$})$, $\mu_{s_os_n}$ and $\gamma_{is_os_n}$ are fixed and patient-level random effects governing patient $i$'s baseline transition probability from stage $s_o$ to $s_n$, and $\tau_{s_os_n}$ and $\alpha_{is_os_n}$ are fixed and patient-level random effects governing the shift from that baseline probability when patient $i$ has an OSA event in the current epoch. Setting $\delta_{ijs_os_n}=0$ for $s_o=s_n$, we specify the current stage as the reference level so that the parameters of the fixed and random effects can be interpreted as positive or negative changes relative to the probability of staying in stage $s_o$. We will return shortly to the implied vector of random effect parameters for each patient, $(\gamma_{iRA}, \gamma_{iRN}, \gamma_{iNA}, \gamma_{iNR}, \alpha_{iRA}, \alpha_{iRN}, \alpha_{iNA}, \alpha_{iNR})^T \in \mathbb{R}^8$, where, for conciseness, we use subscripts A, R, and N to refer to the stages of awakeness, REM, and non-REM sleep.

To facilitate computation, we express the above model as a multinomial distribution for the associated transition counts. In particular, we let $c_{ihs_o} \sim \text{Mult}(\pi_{ihs_oA}, \pi_{ihs_oR}, \pi_{ihs_oN}, n_{ihs_o})$
for $s_o \in \{\text{REM, non-REM}\}$, where $c_{ihs_o} \in \mathbb{Z}^{3}_+$ is the vector of counts of transitions from stage $s_o$ to each of the stages Awake, REM, and non-REM for patient $i$ during epochs for which $v_{ij}=h$. In the above, $n_{ihs_o}$ is the number of epochs the patient $i$ spends in the sleep stage $s_o$ with the status of the event $h \in \{0,1\}$, excluding the final recorded epoch and $\pi_{ihs_os_n} = Pr(s_{i,j+1} = s_n | s_{ij} = s_o, v_{ij} = h)$.

Jointly with the above, we model the inter-event times between apnea/hypopnea events (between the end of one event and the start of the next) using a mixed effects Poisson process for each sleep stage (REM and non-REM). Letting $w_{ils}$ be the $l$th inter-event time for patient $i$ during the sleep stage $s$, we let
$w_{ils} \sim \text{Exp}(\lambda_s \exp(\phi_{is}))$ where $\lambda_s$ captures the overall rate of events in sleep stage $s$ for an average patient, while $\phi_{is}$ is a patient-level random effect capturing variability away from this rate in sleep stage $s$. Similarly to our collapsing the sleep-stage transitions into multinomial counts, here we can express the likelihood for the inter-event times in terms of a Poisson distribution for the counts of events in each stage. That is,
$v_{is} \sim \text{Poisson}(t_{is} \lambda_s \exp(\phi_{is}))$
where $v_{is}$ is the number of events patient $i$ experiences in sleep stage $s$, and $t_{is}$ is the time in seconds patient $i$ spends in sleep stage $s$, excluding intervals when they were actively having an event. Given the above, we now have defined ten random effects for each patient:
\begin{align*}
    \theta_i = (\gamma_i, \alpha_i, \phi_i)^T 
    = (\gamma_{iRA}, \gamma_{iRN}, \gamma_{iNA}, \gamma_{iNR}, \alpha_{iRA}, \alpha_{iRN}, \alpha_{iNA}, \alpha_{iNR}, \phi_{iR}, \phi_{iN})^T \in \mathbb{R}^{10}.
\end{align*}
We assign the vector $\theta_i$ a multivariate normal distribution
$\theta_i \sim N_{10}(0, \Sigma),$
where we learn $\Sigma$ using a latent factor model. That is, we let
$\theta_i = \Lambda \eta_i + \varepsilon_i,$
$\eta_i \sim N_q(0, I),$ $\varepsilon_i \sim N_{10}(0, \Omega)$,
where $\Lambda \in \mathbb{R}^{10 \times q}$ is an unknown factor loading matrix and $\Omega = \text{diag}(\omega_1^2,...,\omega_{10}^2)$. This induces the covariance model
$\theta_i \sim N_{10}(0, \Lambda \Lambda^T + \Omega).$
By choosing $q < p$, this approach allows us to reduce the number of parameters needed to estimate $\Sigma$ from $p(p+1)/2$ for a general covariance matrix to $p(q+1)$, allowing for greater efficiency in estimation. In addition, while $\Lambda$ and $\eta_i$ are in general not identifiable in the above, using the MatchAlign procedure of \cite{poworoznek2021efficiently} we can examine posterior summaries of the $q$ columns of $\Lambda$, which can be thought of as describing low-dimensional components of $\theta_1,...,\theta_n$, similar to principal component vectors, but without the restriction that they are orthogonal.

\subsection{Prior distributions and implementation details}

To complete our Bayesian model specification, we assign prior distributions for all parameters. We give $\text{Gamma(1,10)}$ priors to the $\lambda_s$s, implying a prior mean of $0.1$ with increasing density approaching $0$; this places most prior mass on reasonable rate values, since $1/\lambda_s$ represents the expected inter-event time in stage $s$ in seconds. We assign $\text{Normal(0,10)}$ priors to the $\mu_{s_o s_n}$s and $\tau_{s_o s_n}$s corresponding to a weakly informative prior on the scale of the linear predictor of our model. For the factor model covariance, we give $\text{Normal(0,10)}$ priors to the elements of $\Lambda$ and $\text{Inverse-Gamma}(1,1)$ to the diagonal elements of $\Omega$; similar to the above, these choices give the model some regularization, while being only weakly informative and hence allowing the data to drive the posterior. To sample from the posterior distribution, we implemented the model using the probabilistic programming language Stan \citep{carpenter2017stan}. The code implementing our approach can be found in the online supplement.


\subsection{Loss-based clustering by $\theta_i$}

Usual Bayesian clustering relies on mixture models, leading to computational hurdles and sensitivity to kernel misspecification \citep{rigon2023generalized, dombowsky2025bayesian}. In contrast, we propose a Bayesian decision theoretic approach for inferring clinically interesting subgroups based on the posterior distribution of the patient random effects $\theta_1,\ldots,\theta_n$. The empirical distribution of these random effects can differ arbitrarily from their initial population distribution due to the abundant information we obtain on each patient through the PSG. We focus on estimating the partition $C=\{C_1,\ldots,C_K\}$ of patient indices $\{1,\ldots,n\}$ into $K$ clusters. This is equivalent to choosing a vector of cluster labels $c=(c_1,\ldots,c_n)$ where each $c_i \in \{1,\ldots,K\}$. As a loss function quantifying inconsistency between a candidate clustering and a particular random effects vector, we choose the K-means loss. The Bayes optimal clustering is then chosen to minimize the expected loss marginalizing over the posterior distribution for the random effects. In Section \ref{cluster_point_est} we provide algorithmic details, while in Section \ref{cluster_UQ} we describe an approach for characterizing uncertainty in clustering and for propagating such uncertainty to subsequent analysis results.

\subsubsection{Point estimation}\label{cluster_point_est}

Accounting for posterior uncertainty in inferring the random effects for each patient, our Bayesian estimator of the cluster labels and centers corresponds to minimizing the 
expected K-means loss over $\pi_{post}$:
\begin{align}\label{opt_joint}
    \left(\hat{c}, \hat{b}\right) = \argmin_{c: |C|=K, \,\, b \in \mathbb{R}^{K \times d}} \sum_{k=1}^K \sum_{i \in C_k} \mathbb{E}_{\pi_{post}} \left[||\theta_i - b_k||^2 \right].
\end{align}
When applying K-means clustering to observed data rather than uncertain parameters, the problem is often framed as optimizing only over $c$, with centers $b_1,...,b_K$ automatically defined as the means of observations in clusters $1,...,K$. Adapting such a formulation to our case produces 
\begin{align}\label{opt_simple}
    \hat{c}' = \argmin_{c: |C|=K} \sum_{k=1}^K \sum_{i \in C_k} \mathbb{E}_{\pi_{post}} \left[||\theta_i - \overline{\theta}_k||^2 \right],
\end{align}
where each cluster center $\overline{\theta}_k$ is a random variable that varies over posterior samples. Note that \eqref{opt_joint} and \eqref{opt_simple} are not equivalent and, in general, the learned partitions defined by $\hat{c}$ and $\hat{c}'$ will differ. In the results that follow, we focus on the formulation \eqref{opt_joint}; we are interested in interpreting the cluster summaries in the context of the estimated cluster centers, so it is appealing to obtain a principled point estimate of these centers jointly with the assignments. If we first compute $\hat{c}'$ using \eqref{opt_simple} and then summarize the centers $\hat{b}'$ as posterior means conditional on $\hat{c}'$, we produce a suboptimal solution $(\hat{c}', \hat{b}')$ with respect to \eqref{opt_joint}. However, we also implement formulation \eqref{opt_simple} for comparison and summarize the results in Supplementary Information.

In addition to its interpretability, the formulation \eqref{opt_joint} is easy to compute, as implied by Proposition \ref{proposition_1}.

\begin{proposition}\label{proposition_1}
Let $(\hat{c}, \hat{b})$ be the solution to \eqref{opt_joint}. Then we can equivalently compute
$$\left(\hat{c}, \hat{b}\right) = \argmin_{c: |C|=K, \,\, b \in \mathbb{R}^{K \times d}} \sum_{k=1}^K \sum_{i \in C_k} ||\mathbb{E}_{\pi_{post}}[\theta_i] - b_k||^2.$$
That is, we can simply apply the standard K-means algorithm to the posterior means of $\theta_1,...,\theta_n$.
\end{proposition}
The proof of Proposition \ref{proposition_1} is straightforward and is given in the Supplemental Information. With the efficiency of modern implementations of K-means, such as the \lstinline{kmeans} function in R \citep{R_lang}, this result suggests that finding the Bayes-optimal clustering is computationally trivial. By choosing the number of posterior samples to be large, we can estimate $\mathbb{E}_{\pi_{post}}[\theta_i]$ arbitrarily well with Monte Carlo sample averages.

\subsubsection{Uncertainty quantification}\label{cluster_UQ}

To characterize uncertainty in clustering patients into groups, we focus on the conditional posterior probability of assigning patient $i$ to cluster $k$ treating the cluster centers as fixed at $\hat{b}$: 
\begin{align}\label{uq_prob}
    Pr(c_i = k | b = \hat{b}) = \mathbb{E}_{\pi_{post}} \left[ \mathds{1} \left(\argmin_{k' \in \{1,...,K\}} ||\theta_i - \hat{b}_{k'}||^2 = k \right) \right].
\end{align}
A Monte Carlo estimate of the probabilities in \eqref{uq_prob} can be easily calculated with posterior samples, and these probabilities can then be propagated to account for the uncertainty in future patient summaries based on cluster assignment.

For interpretability, we purposely avoid accounting for uncertainty in inferring the cluster centers. Fixing the centers gives a clear interpretation of statements such as ``patient $i$ is in cluster $k$ with probability $p$'' that would be muddled if the location of the cluster $k$ was also in question. In addition, 
we are interested in summarizing clinically interesting groups of our $n$ patients without assuming true, well-separated patient clusters. Given this goal, it is reasonable to fix the locations of these group of patients according to \eqref{opt_joint} and then examine the uncertainty relative to these locations. Thus, our results in Section \ref{section_application} account only for the uncertainty in \eqref{uq_prob}. Two alternative approaches are described and implemented in the Supplementary Information.

\section{Simulations}\label{section_simulations}

To evaluate our approach, we simulated PSG data in three scenarios. In Scenario 1, we generated data from our model as specified. For each replication, we drew $\mu_{s_o s_n}$ parameters independently from a $\text{Uniform}[-4,-2]$ distribution, $\tau_{s_os_n}$ from $\text{Uniform}[0,1]$, $\lambda_s$ from $\text{Uniform}[0.01,0.03]$, and two additional parameters $\mu_{AR}$ and $\mu_{AN}$ that govern transitions from Awake from $\text{Uniform}[-3,-1]$. We generate a factor model covariance for random effects with $q=3$ factors. Factor loadings were drawn independently from $N(0,0.2)$ distributions and idiosyncratic variances were set to $0.2$. Random effects $\theta_i$ were drawn from the resulting multivariate normal distribution. 
In Scenario 2, we used the same setup, except that $\mu_{s_os_n}$ and $\lambda_s$ varied over time, so that the baseline rates of stage transitions and OSA events were not constant during the night. Such non-stationarity is likely to occur in practice, and it is important to verify our inferences are robust to this misspecification.
We drew $\mu_{s_os_n}$ as linear functions with intercepts from $\text{Uniform}[-4,-3]$ and slopes from $\text{Uniform}[-0.001,0.001]$ in units of epochs, so that over a 1000-epoch night, a parameter could shift by up to one unit. We drew $\lambda_s$ intercepts from $\text{Uniform}[0.01,0.03]$ and slopes from $\text{Uniform}[-0.00001,0.00001]$, for a total shift of up to $0.01$. In Scenario 3, we generate fixed effects as in Scenario 1, but drew random effects $\theta_i$ from a mixture of four normals with means drawn independently from $N_{10}(0, 0.25I)$. Patients were assigned uniformly at random to one of the four groups and their $\theta_i$ was drawn from a multivariate Gaussian with the appropriate mean and variance $0.25I$. For identifiability, we  centered $\theta_i$ values at mean zero. For each scenario and each of $100$ random initializations, we generated one night of data for each of $1000$ patients, with each night recorded for $1000$ epochs including any awake intervals; this corresponds to eight hours and twenty minutes of recording for each patient.

\begin{table}[b]
\centering
\caption{\small{Accuracy of parameter estimates under three data generation scenarios. In scenario 1 the model is correctly specified, while in scenarios 2 and 3 there are different types of misspecification.}}
\begin{tabular}{llll}
\hline
                                               & Scenario 1 & Scenario 2 & Scenario 3 \\ \hline
Fixed Effect MSE                               & 0.00101    & N/A        & 0.000482           \\
Fixed Effect Coverage                          & 0.943      & N/A        & 0.985           \\
Random Effect Covariance MSE                   & 0.00159     & 0.00163     & N/A           \\
Random Effect Covariance Coverage              & 0.945      & 0.953      & N/A           \\
Random Effect MSE                              & 0.107      & 0.120      & 0.0993           \\
Random Effect Coverage                         & 0.949      & 0.948      & 0.951                    
\end{tabular}
\label{table_sim_results}
\end{table}

A summary of our model's accuracy is shown in Table \ref{table_sim_results}, averaged over $100$ replications. Accuracy is similar across the three scenarios. Coverage of $95\%$ credible intervals is excellent for fixed effects, random effects, and elements of the random effect covariance matrix across the board, with only slight differences between the different data-generating processes. For the random effects and their covariance, which are focal points of our applied results, mean squared error does not suffer substantially given non-constant fixed effects in Scenario 2, and the covariance $95\%$ interval coverage actually improves slightly compared to Scenario 1 where our model is correctly specified. This is highly encouraging that our model performs well in capturing patient heterogeneity, even when our assumptions of constant sleep dynamics and event rates are substantial misspecifications. It is somewhat more difficult to interpret comparisons in random effect MSE for Scenario 3, since the random effects are drawn from a mixture that likely has a somewhat different scale than the factor model covariance, but in general performance is good, with the smallest MSE and excellent coverage for the random effects, as well as smaller MSE and improved coverage for the fixed effects, which are generated identically to those in Scenario 1. Overall, these results suggest that our inferences are reasonably robust to misspecification, both in terms of non-constant dynamics and event rates, and in terms of substantial misspecification of the random effects distribution.

Finally, to evaluate the efficacy of our clustering approach, for Scenario 3 data, we additionally generated an outcome variable for each patient from a mixture distribution with the same assignments as the mixture for $\theta_i$. The means for these components were chosen without replacement from $\{1,2,3,4\}$, and then the patients' values were drawn with Gaussian noise with standard deviation $0.2$. We then clustered patients in two ways, first using our approach described in Section \ref{cluster_point_est}, and then for comparison, using K-means with standardized values of REM AHI, non-REM AHI, total time in REM, and total time in non-REM as input data. These represent simple summary statistics, often used to infer stage-specific event rates and sleep architecture. Table \ref{table_sim_clustering} compares the adjusted Rand index for each clustering compared to the true mixture components averaged over 100 replications. Our approach has a dramatically better clustering accuracy, with an ARI of $\sim 0.7$, compared to a poor ARI of $\sim 0.2$ for the competitor. The competitor finds bigger clusters on average, as the within-cluster outcome variance is over double that for our approach. 

\begin{table}[t]
\centering
\caption{\small{Accuracy of clustering based on $\hat{\theta}_i$ compared to using patient summaries of REM AHI, non-REM AHI, time in REM, and time in non-REM, evaluated on Scenario 3 simulated data.}}
\begin{tabular}{lll}
\hline
                         & Our approach & K-means with basic summaries \\ \hline
Adjusted Rand Index & 0.670                              & 0.229                             \\
Within-cluster outcome variance             & 0.447                              & 0.959                            
\end{tabular}
\label{table_sim_clustering}
\end{table}

\section{Application}\label{section_application}

\subsection{Exploratory data analysis}
\label{EDA}

Before fitting our model, we performed some data exploration, summarized in Table \ref{table_EDA}. Observe that average AHI appears to be somewhat higher during REM sleep compared to non-REM sleep. Interestingly, event durations also appear to be longer in REM vs. non-REM. This not only adds to the evidence that OSA intensity may be higher for the average person in REM sleep compared to non-REM, but actually suggests that the difference between AHI values in REM vs. non-REM may in some sense understate the higher rate during REM sleep, since longer average events lead to a lower AHI given the same distribution of times between events. (E.g., a patient with two 10-second events each minute will have double the AHI of a patient with one 40-second event each minute, even though they both have an average inter-event time of 20 seconds, and more of the second patient's sleep time is spent having events.) This suggests our approach of modeling inter-event times as the time between the end of one event and the start of the next may more reasonably capture the difference in in OSA intensity compared to AHI. 

\begin{table}[t]
\centering
\caption{\small{Patient means and standard deviations for stage-specific AHI, event durations, and time spent in each sleep stage.}}
\resizebox{\textwidth}{!}{%
\begin{tabular}{llll}
\hline
        & AHI         & Mean event duration (seconds) & Time spent in stage (hours) \\ \hline
REM     & 49.6 (22.7) & 27.3 (11.4)                   & 1.1 (0.5)                   \\
non-REM & 43.5 (24.4) & 24.1 (6.7)                    & 5.2 (0.9)                   \\ \hline
\end{tabular}}
\label{table_EDA}
\end{table}

Finally, the average patient spends far more time in non-REM than REM sleep. This suggests that sleep stage-specific event rates can affect sleep quality and health through different pathways. On one hand, since non-REM represents the majority of sleep, an elevated rate of non-REM events has a larger impact on the total number of events in a night than an elevated REM rate. However, since REM sleep is believed to be particularly important for learning and memory \citep{peever2017biology}, if an elevated rate of REM events causes a patient's small but important amount of REM sleep to be interrupted or cut short, this could also be highly detrimental. Box plots showing distributions of the quantities in Table \ref{table_EDA} are included in the Supplementary Information.

\subsection{Model fitting}

To sample from the posterior, we implemented our model in the probabilistic programming language Stan \citep{carpenter2017stan}. We ran four chains, each for 2500 burn-in iterations and 2500 subsequent posterior samples, for a total of 10,000 posterior samples. To choose the number of factors $q$ in the factor model for random effects, we ran a principal component analysis on the estimated random effect vectors in a simpler model with diagonal covariance, and examined the scree plot; this resulted in a choice of $q=5$. The mixing appears satisfactory, with all effective sample sizes larger than 2700 and all Gelman-Rubin diagnostic values less than 1.02 in all main effects, random effects, and random effect covariance terms.

\begin{figure}[b]
\includegraphics[width=\textwidth]{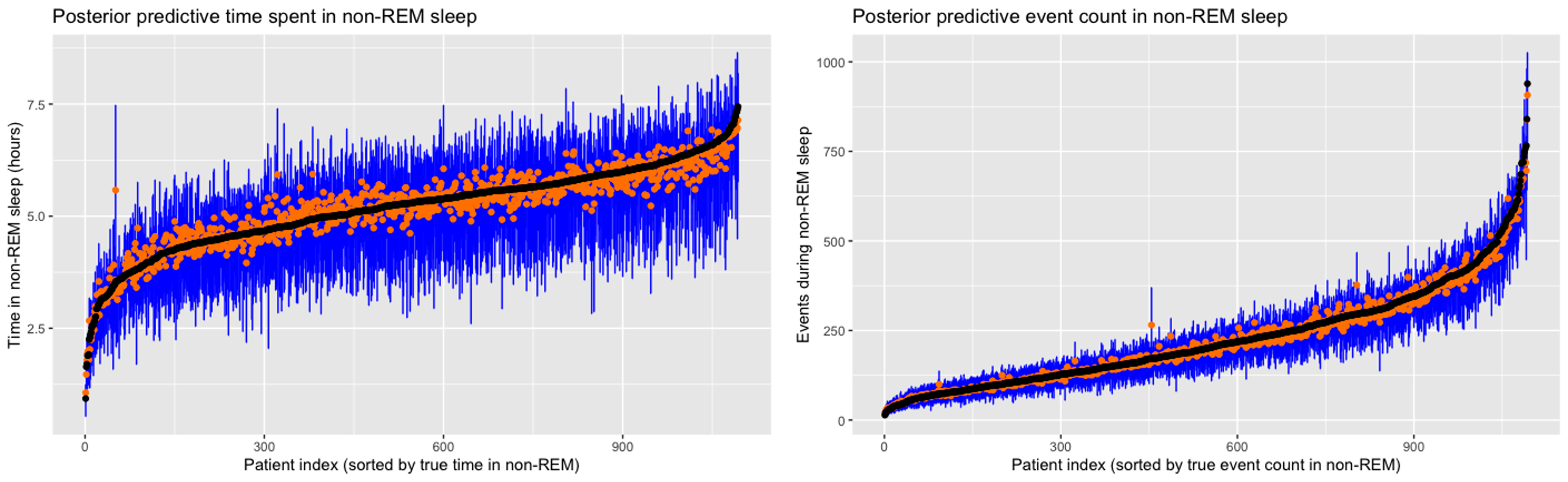}
\caption{\small{Posterior predictive distributions of time spent in non-REM sleep (left) and the number of events occurring during non-REM sleep (right). Black dots are observed values; orange dots are posterior predictive means; blue lines are $95\%$ posterior predictive intervals.}}
\label{post_pred_nonREM}
\end{figure}

To evaluate the quality of model fit, we performed several posterior predictive checks. To simulate a night's data for each patient and each posterior sample, we simulated sleep stage transitions and OSA event occurrences conditional on the person's total number of epochs asleep. If a person transitioned to Awake, we draw the next sleep stage uniformly at random from that patient's observed initial sleep stages after waking up. To generate the duration of events, we draw the duration of events uniformly at random from the patient's list of event durations for the current sleep stage. Figure \ref{post_pred_nonREM} shows the posterior predictive means (orange dots) and $95\%$ intervals (blue lines) for the amount of time spent in non-REM sleep and the number of total OSA events occurring during non-REM sleep. The black dots indicate the observed value for each patient. The intervals contain the observed values for $99.9\%$ of the patients for time spent in non-REM and $99.8\%$ of the patients for the count of non-REM events. The results are similar for REM sleep, with $95\%$ interval coverage of $99.9\%$ and $98.9\%$. The posterior predictive plots for time and event count in REM as well as posterior predictive AHI values in REM and non-REM are presented in the Supplementary Information.

\subsection{Posterior summaries}

\subsubsection{Model parameters}

\begin{figure}[b]
\centering
\includegraphics[width=\textwidth]{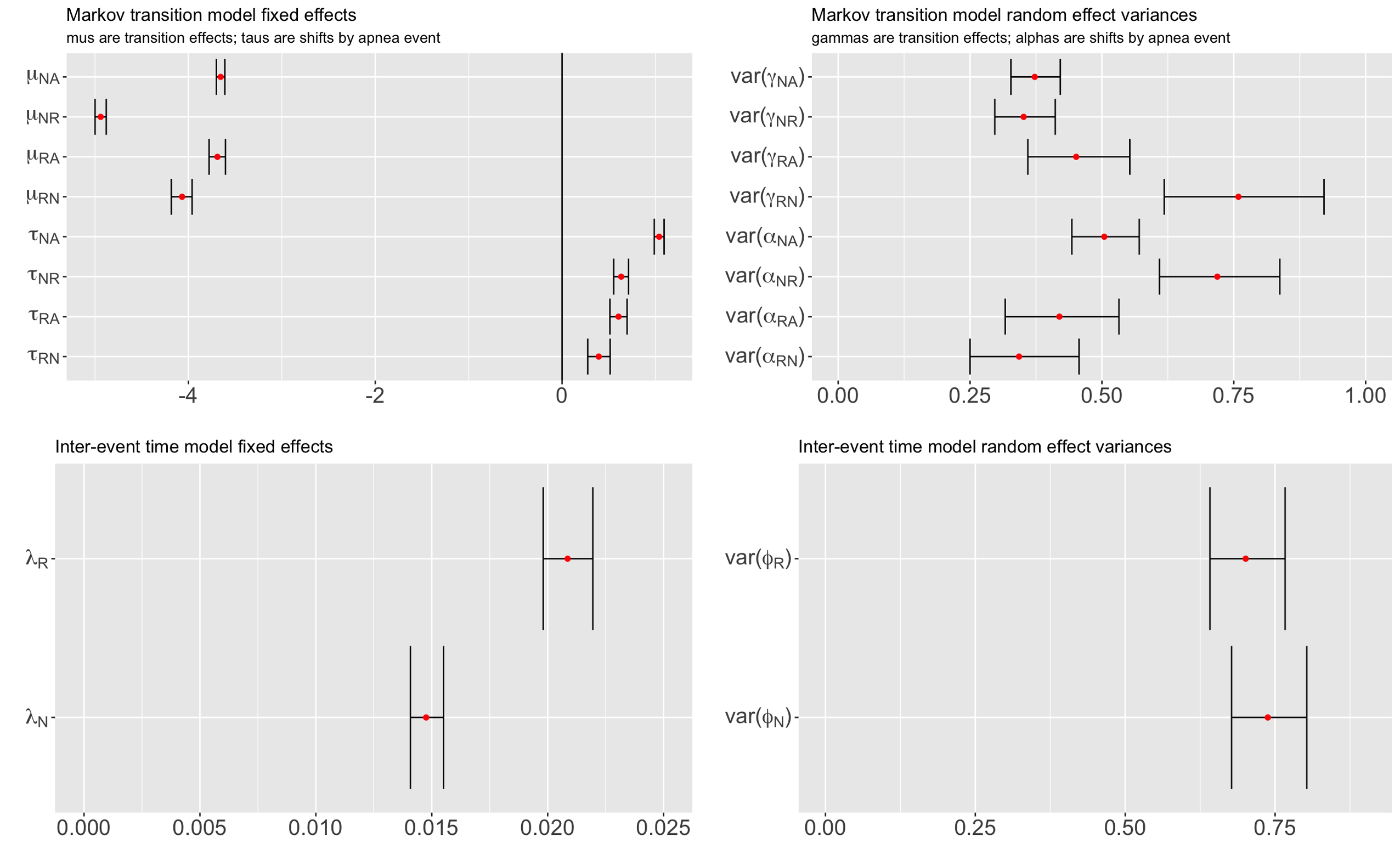}
\caption{\small{Posterior summaries of the Markov model fixed effects (top left) and random effect variances (top right), and the apnea and hypopnea inter-event time model fixed effects (bottom left) and random effect variances (bottom right). Dots represent posterior means, and intervals represent 95\% posterior credible intervals.}}
\label{fig_fixed_random_effects_4panel}
\end{figure}

Figure \ref{fig_fixed_random_effects_4panel} summarizes the posterior distributions for the transition and inter-event model parameters. In the top left panel, observe that the posteriors of all $\mu_{s_os_n}$ parameters are strongly negative, indicating that for both REM and non-REM sleep, patients are overwhelmingly likely to stay in the same stage from one epoch to the next. The 95\% posterior credible intervals for $\tau_{RN}$, $\tau_{RA}$, $\tau_{NR}$, and $\tau_{NA}$ are all strictly positive, indicating that having an apnea event in a given epoch makes the average patient more likely to transition out of their current sleep stage. The $\tau_{RA}$ and $\tau_{NA}$ have magnitudes larger than $\tau_{RN}$ and $\tau_{NR}$, respectively, indicating that apnea events have a particularly strong effect on the probability of awakening. In the upper right panel of Figure \ref{fig_fixed_random_effects_4panel}, the baseline transition parameters $\gamma_{is_os_n}$ appear to have a larger variance for transitions out of REM than for transitions out of non-REM, while the opposite is true for the parameters $\alpha_{is_os_n}$. This suggests that there may be more baseline patient variation in transitions from REM sleep, but more variation in the effect of OSA events during non-REM sleep.

As expected from our EDA in Section \ref{EDA}, the bottom left panel of Figure \ref{fig_fixed_random_effects_4panel} shows that the baseline rate of events is substantially higher during REM sleep compared to non-REM. The random effect variances shown in the bottom right panel are very similar between REM and non-REM, with the posterior mean slightly higher for non-REM. However, since these random effects $\phi_{is}$ affect the expected inter-event times at the patient level multiplicatively, the actual patient-level variability is higher during REM sleep than during non-REM, due to the higher baseline rate during REM. Finally, given the setup of the model in Section \ref{Model_description}, the variances observed for the $\phi_{is}$s indicate substantial patient variation in the time between events. If the $\phi_{is}$ variances are $\sim 0.7$, this corresponds to a patient with a rate one standard deviation above the mean having an expected inter-event time over five times shorter than a patient whose rate is one standard deviation below the mean. 

\begin{figure}[t]
\centering
\includegraphics[width=0.7\textwidth]{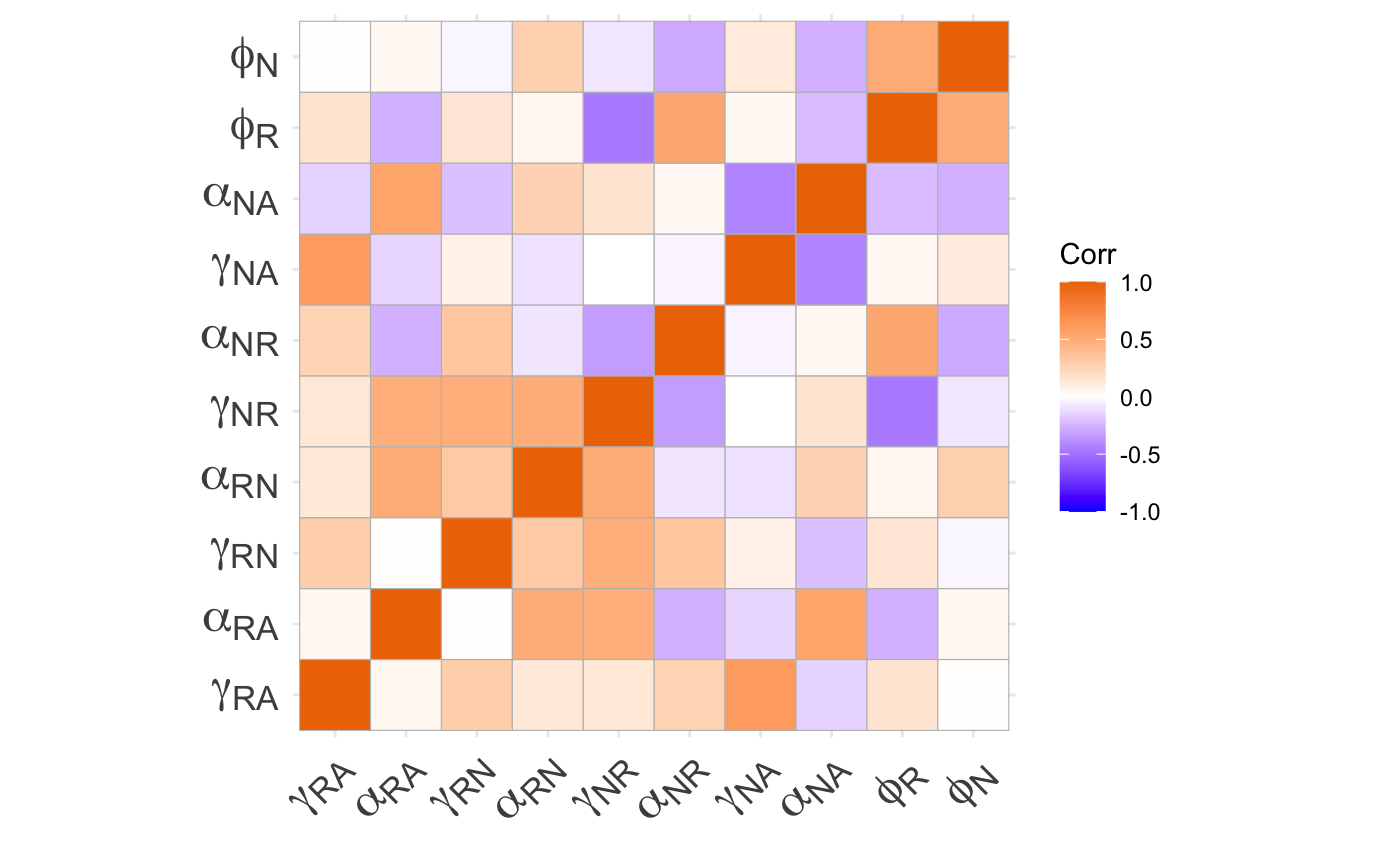}
\caption{\small{Posterior mean correlation matrix for the random effect vectors $\theta_i$.}}
\label{fig_corrplot}
\end{figure}

To illustrate the relationships among random effects at the patient level, Figure \ref{fig_corrplot} shows the posterior mean of the correlation matrix for $\theta_i$, which provides a number of insights into the patient distribution. First, most correlations between the dynamics parameters $\gamma$ and $\alpha$ are positive, particularly for parameters controlling the transitions from REM sleep, suggesting that individuals likely to transition from REM sleep at baseline are also more affected than average by OSA events during REM. However, for non-REM parameters, the negative correlations between $\gamma_{NR}$ and $\alpha_{NR}$, and between $\gamma_{NA}$ and $\alpha_{NA}$, suggest that patients with high baseline rates of transitions from non-REM sleep tend to experience a smaller disruption than average due to non-REM events. This may be due in part to the logit scale in our model, which implies that the interpretation of $\alpha_{NA}$ and $\alpha_{NR}$ depends on the values of $\gamma_{NA}$ and $\gamma_{NR}$. Finally, as expected, the random effects between events $\phi_R$ and $\phi_N$ are positively correlated, suggesting that patients with a high rate of OSA events in REM or non-REM sleep are more likely to also have a high rate in the other. The interpretation of the relationship between these parameters $\phi$ and the dynamic random effects is more subtle, with a number of moderate positive and negative correlations. However, it is interesting to observe the negative correlations between $\phi_R$ and $\alpha_{RA}$, and between $\phi_N$ and $\alpha_{NA}$, which suggest that patients with a higher rate of OSA events may tend to have their sleep less disrupted by each event.

In the online supplement, we summarize the posterior means of the factor loadings in the matrix $\Lambda$, aligned for interpretability using the method of \cite{poworoznek2021efficiently}. In general, the interpretations of these factors yield similar insights to examining the correlations in Figure \ref{fig_corrplot}. Factors 1 and 5 yield the positive correlations among the REM transition parameters, factor 4 yields the positive correlation between $\phi_R$ and $\phi_N$, and factors 2 and 3 are primarily responsible for the negative correlations discussed above.

\subsubsection{Clustering}

Given the posterior samples of $\theta_i$, we computed point estimates of the cluster assignments and centers based on \eqref{opt_joint} using the \lstinline{kmeans} function in R \citep{R_lang}, as well as the associated assignment probabilities of \eqref{uq_prob}. Letting the number of clusters $K=4$ based on an elbow plot for the within-cluster sum of squares, the estimated centers are shown in Table \ref{table_cluster_centers}. One way to interpret these centers in terms of sleep and OSA characteristics is to consider whether the $\gamma$ and $\alpha$ parameters tend to be high or low, suggesting highly disrupted vs. steady sleep, and whether each of $\phi_R$ and $\phi_N$ are high or low. Using this framework, we find that cluster 1 (276 patients) has moderate sleep dynamics, a low event rate in REM sleep, and a moderate event rate in non-REM sleep, cluster 2 (143 patients) has highly disrupted sleep dynamics with moderate REM and non-REM event rates, cluster 3 (326 patients) has fairly calm sleep dynamics, a moderate event rate in REM, and a low event rate in non-REM, and cluster 4 (348 patients) has fairly calm sleep dynamics but high event rates in both REM and non-REM. 

\begin{table}[t]
\centering
\caption{\small{Bayes-optimal cluster center estimates based on k-means loss with $k=4$.}}
\resizebox{0.85\textwidth}{!}{%
\begin{tabular}{rrrrrrrrrrr}
  \hline
 & $\gamma_{RA}$ & $\gamma_{NA}$ & $\alpha_{RA}$ & $\alpha_{NA}$ & $\gamma_{RN}$ & $\gamma_{NR}$ & $\alpha_{RN}$ & $\alpha_{NR}$ & $\phi_R$ & $\phi_N$ \\
  \hline
Cluster 1 & -0.17 & 0.00 & 0.30 & 0.17 & -0.37 & 0.31 & 0.02 & -0.86 & -0.97 & -0.14 \\ 
  Cluster 2 & 0.62 & 0.18 & 0.38 & 0.08 & 1.29 & 0.64 & 0.61 & 0.53 & 0.18 & -0.03 \\ 
  Cluster 3 & -0.11 & -0.19 & -0.20 & 0.16 & -0.09 & -0.21 & -0.26 & 0.51 & -0.01 & -0.73 \\ 
  Cluster 4 & -0.02 & 0.10 & -0.20 & -0.31 & -0.16 & -0.31 & -0.02 & -0.02 & 0.69 & 0.79 \\ 
   \hline
\end{tabular}}
\label{table_cluster_centers}
\end{table}

\begin{figure}[b]
\centering
\includegraphics[width=\textwidth]{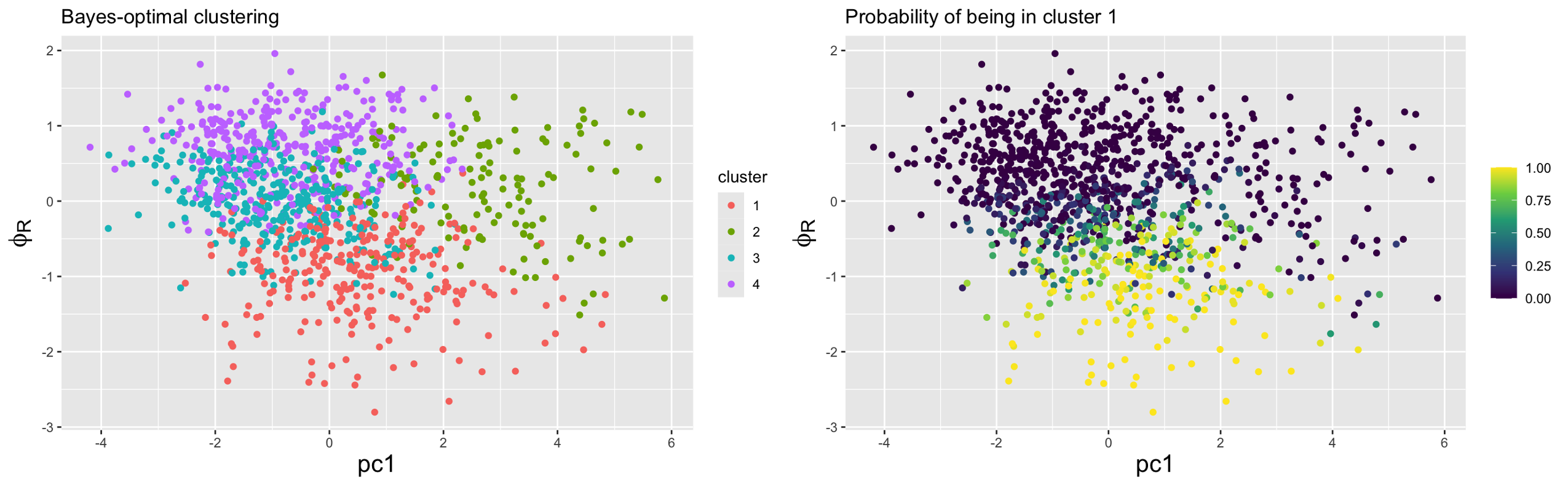}
\caption{\small{Bayes-optimal point estimate of cluster assignments based on expected K-means loss for random effect vectors $\theta_i$ (left), and posterior probability of assignment to cluster 1, computed with equation \eqref{uq_prob} (right). pc1 refers to the first principal component vector computed with the posterior means of the dynamics model random effects.}}
\label{fig_cluster_2panel}
\end{figure}

The cluster assignment point estimates are shown in the left panel of Figure \ref{fig_cluster_2panel}, indicated by colored points showing each $\theta_i$ projected into two dimensions. The vertical axis is simply $\phi_R$, while the horizontal axis is the first principal component score for PCA performed on the posterior means of the sleep-dynamics random effects, $(\gamma_{RA},\allowbreak \gamma_{NA}, \allowbreak \alpha_{RA}, \allowbreak \alpha_{NA}, \allowbreak \gamma_{RN}, \allowbreak \gamma_{NR}, \allowbreak \alpha_{RN}, \allowbreak \alpha_{NR})$. The vector of the first associated principal component is $(0.28, \allowbreak 0.06, \allowbreak 0.48, \allowbreak 0.23, \allowbreak 0.33, \allowbreak 0.50, \allowbreak 0.52, \allowbreak -0.07)$, suggesting that a positive value of pc1 roughly corresponds to more disruption of the sleep stages, both at baseline and due to OSA events. Given this interpretation, the projected locations of the clusters appear to correspond to the above interpretation of their centers.

To illustrate the uncertainty quantification produced by equation \eqref{uq_prob}, the right panel of Figure \ref{fig_cluster_2panel} shows the probabilities of assignment to cluster 1. Although the majority of patients have probabilities very close to 0 or 1, a number of observations near the boundary with other clusters have meaningful levels of uncertainty in their assignment.

\subsection{Relationship of cluster assignment with other variables}

Given the cluster assignments learned above, we can examine how other variables of interest vary by cluster. Table \ref{table_demographics} shows the weighted means and standard deviations of a number of demographic and clinical variables by the probability of cluster assignment. Several demographic variables appear to vary meaningfully between groups. For example, cluster 1 has a substantially higher percentage of males than the other groups, as well as the highest percentage of white patients. Examining other health variables, cluster 4 has an extremely high average BMI of 35.2, as well as an AHI of 63.7, both of which are substantially higher than those of clusters 1-3. The total amount of sleep appears reasonably similar across clusters, but the portion of that sleep spent in REM vs. non-REM differs substantially. In particular, cluster 2 has well under an hour of REM sleep on average, aligning with the highly disturbed sleep indicated by the cluster 2 center in Table \ref{table_cluster_centers}. Finally, Table \ref{table_demographics} shows the average scores for Pathfinder and BSRT for each group. Although these generally appear fairly similar, clusters 2 and 4 have the worst two average scores on both tests.

\begin{table}[t]
\centering
\caption{\small{Weighted means and standard deviations of demographic and PSG variables by cluster assignment. The weights used are the posterior probabilities of assignment to each cluster.}}
\begin{tabular}{lllll}
\hline
                       & Cluster 1   & Cluster 2   & Cluster 3   & Cluster 4   \\ \hline
Age                    & 51.8 (13.3) & 52.7 (11.1) & 50.5 (12.2) & 51.9 (11.8) \\
Fraction male          & 0.79        & 0.64        & 0.54        & 0.66        \\
Fraction white         & 0.82        & 0.77        & 0.73        & 0.74        \\
BMI                    & 29.8 (5.4)  & 31.2 (6.2)  & 31.5 (6.8)  & 35.2 (7.9)  \\
AHI                    & 31.2 (17.4) & 38.0 (22.1) & 23.4 (10.2) & 63.7 (23.7) \\
Hours asleep           & 6.26 (1.08) & 6.26 (1.02) & 6.43 (1.07) & 6.28 (1.11) \\
Hours in REM sleep     & 1.18 (0.47) & 0.86 (0.44) & 1.31 (0.48) & 1.04 (0.52) \\
Hours in non-REM sleep & 5.07 (0.90) & 5.41 (0.90) & 5.12 (0.87) & 5.25 (0.97) \\
Pathfinder             & 24.1 (7.0)  & 24.4 (5.8)  & 24.3 (6.2)  & 25.0 (7.1)  \\
BSRT                   & 50.2 (9.2)  & 49.0 (9.1)  & 50.4 (9.0)  & 49.2 (9.3) 
\end{tabular}
\label{table_demographics}
\end{table}

\begin{table}[b]
\centering
\caption{\small{Results from Bayesian linear regression models for BSRT and Pathfinder scores on cluster with reference level 4 (model 1) and AHI (model 2), both after accounting for age and sex. For BSRT, the $95\%$ credible interval for the difference between clusters 1 and 4 is strictly positive in model 1, while the $95\%$ credible interval for AHI includes zero in model 2. For Pathfinder, the $95\%$ credible intervals for the cluster differences and for AHI all include zero.}}
\begin{tabular}{llll}
\hline
                                                                                    & Coefficient & Posterior Mean & 95\% Posterior Credible Interval \\ \hline
\multirow{6}{*}{\begin{tabular}[c]{@{}l@{}}BSRT:\\ Regression 1\end{tabular}}       & Intercept   & \textbf{62.76} & \textbf{(60.35, 65.25)}          \\
                                                                                    & Cluster1    & \textbf{1.55}  & \textbf{(0.18, 2.95)}            \\
                                                                                    & Cluster2    & -0.51          & (-2.15, 1.13)                    \\
                                                                                    & Cluster3    & 0.37           & (-0.88, 1.67)                    \\
                                                                                    & Age         & \textbf{-0.21} & \textbf{(-0.25, -0.17)}          \\
                                                                                    & SexM        & \textbf{-4.04} & \textbf{(-5.11, -2.96)}          \\ \hline
\multirow{4}{*}{\begin{tabular}[c]{@{}l@{}}BSRT:\\ Regression 2\end{tabular}}       & Intercept   & \textbf{63.74} & \textbf{(61.34, 66.14)}          \\
                                                                                    & AHI         & -0.02          & (-0.04, 0.00)                    \\
                                                                                    & Age         & \textbf{-0.21} & \textbf{(-0.25, -0.17)}          \\
                                                                                    & SexM        & \textbf{-3.74} & \textbf{(-4.80, -2.63)}          \\ \hline
\multirow{6}{*}{\begin{tabular}[c]{@{}l@{}}Pathfinder:\\ Regression 1\end{tabular}} & Intercept   & \textbf{12.68} & \textbf{(10.93, 14.36)}          \\
                                                                                    & Cluster1    & -0.90          & (-1.80, 0.01)                    \\
                                                                                    & Cluster2    & -1.04          & (-2.15, 0.09)                    \\
                                                                                    & Cluster3    & -0.54          & (-1.39, 0.36)                    \\
                                                                                    & Age         & \textbf{0.24}  & \textbf{(0.21, 0.27)}            \\
                                                                                    & SexM        & 0.21           & (-0.54, 0.94)                    \\ \hline
\multirow{4}{*}{\begin{tabular}[c]{@{}l@{}}Pathfinder:\\ Regression 2\end{tabular}} & Intercept   & \textbf{11.82} & \textbf{(10.08, 13.51)}          \\
                                                                                    & AHI         & 0.01           & (-0.00, 0.03)                    \\
                                                                                    & Age         & \textbf{0.24}  & \textbf{(0.21, 0.27)}            \\
                                                                                    & SexM        & 0.11           & (-0.63, 0.84)                    \\ \hline
\end{tabular}
\label{table_regression_fits}
\end{table}

To illustrate that random effects in our model and the clusters they produce can generate insights missed by a simple AHI approach, Table \ref{table_regression_fits} shows the results of two Bayesian Gaussian linear regression models for each of BSRT and Pathfinder. For each cognitive test, model 1 regresses patient score on cluster label, age, and sex, while model 2 regresses score on AHI, age, and sex. We assign vague $N(0, 100^2)$ priors to the coefficients and an $IG(1,1)$ prior to the residual variance. We fit the models in Stan. The symmetric $95\%$ interval for age is negative for both BSRT models (the score is the total number of words remembered) and positive in both Pathfinder models (the score is the time to complete the task). This suggests that higher age is negatively associated with both verbal memory as measured by BSRT and attention as measured by Pathfinder. The interval for male sex is strictly negative in both BSRT models, but includes zero in both Pathfinder models, suggesting a strong negative association with memory but not attention.

In BSRT model 1, the $95\%$ interval for cluster 1 relative to cluster 4 is strictly positive, with posterior mean estimating a modest $1.55$ point difference in expected score. However, despite the variation in AHI between the groups and the fact that group 4 has by far the highest AHI, in BSRT model 2 the $95\%$ credible interval for AHI contains $0$. While cluster 1 does not have the lowest AHI, it does have the lowest average $\phi_R$, suggesting that the rate of events may matter more in REM sleep than non-REM. Cluster 3 has the lowest AHI and the lowest average $\phi_N$, but the interval for its difference in BSRT performance with cluster 4 includes $0$. Similarly, while cluster 2 also has substantially lower AHI than cluster 4, it also does not differ significantly in BSRT. An interpretation is that the higher AHI of cluster 4 is balanced by the more disturbed sleep of cluster 2, suggesting that examining the occurrences of OSA events and the dynamics of sleep together in our model may be important to understand the drivers of the severity of symptoms. Overall, the possible differential association of cluster labels with memory but not attention is intriguing. REM sleep is implicated in memory consolidation \citep{peever2017biology}, and cluster 1 had the lowest REM event rate. This raises the hypothesis that REM preservation may benefit memory-dependent processes.

\subsection{Alternate clustering based on simple summary statistics}

Examining the AHI and time values in each sleep stage in Table \ref{table_demographics}, it appears that these basic summary statistics vary by cluster. This suggests that despite the lower performance of the simple clustering of these variables in our simulations in Section \ref{section_simulations}, they may also give useful insight here. To explore this, we again grouped patients using the loss function of K-means with $K=4$, but this time using REM AHI, non-REM AHI, time in REM, and time in non-REM as input variables. The estimated cluster centers are shown in Table \ref{table_alternate_cluster_centers}. Cluster 1 has the lowest AHI in each sleep stage and the longest time spent in REM, while cluster 4 has the highest AHI values but the least time spent in REM.

\begin{table}[b]
\centering
\caption{\small{Estimated cluster centers based on K-means loss with REM AHI, non-REM AHI, time in REM, and time in non-REM as input data.}}
\begin{tabular}{lllll}
\hline
          & REM AHI & non-REM AHI & time in REM & time in non-REM \\ \hline
Cluster 1 & 29.2    & 23.9        & 1.22        & 5.07            \\
Cluster 2 & 41.6    & 59.5        & 1.08        & 5.24            \\
Cluster 3 & 64.9    & 36.1        & 1.16        & 5.22            \\
Cluster 4 & 78.0    & 83.3        & 0.90        & 5.32            \\ \hline
\end{tabular}
\label{table_alternate_cluster_centers}
\end{table}

\begin{figure}[t]
\centering
\includegraphics[width=0.8\textwidth]{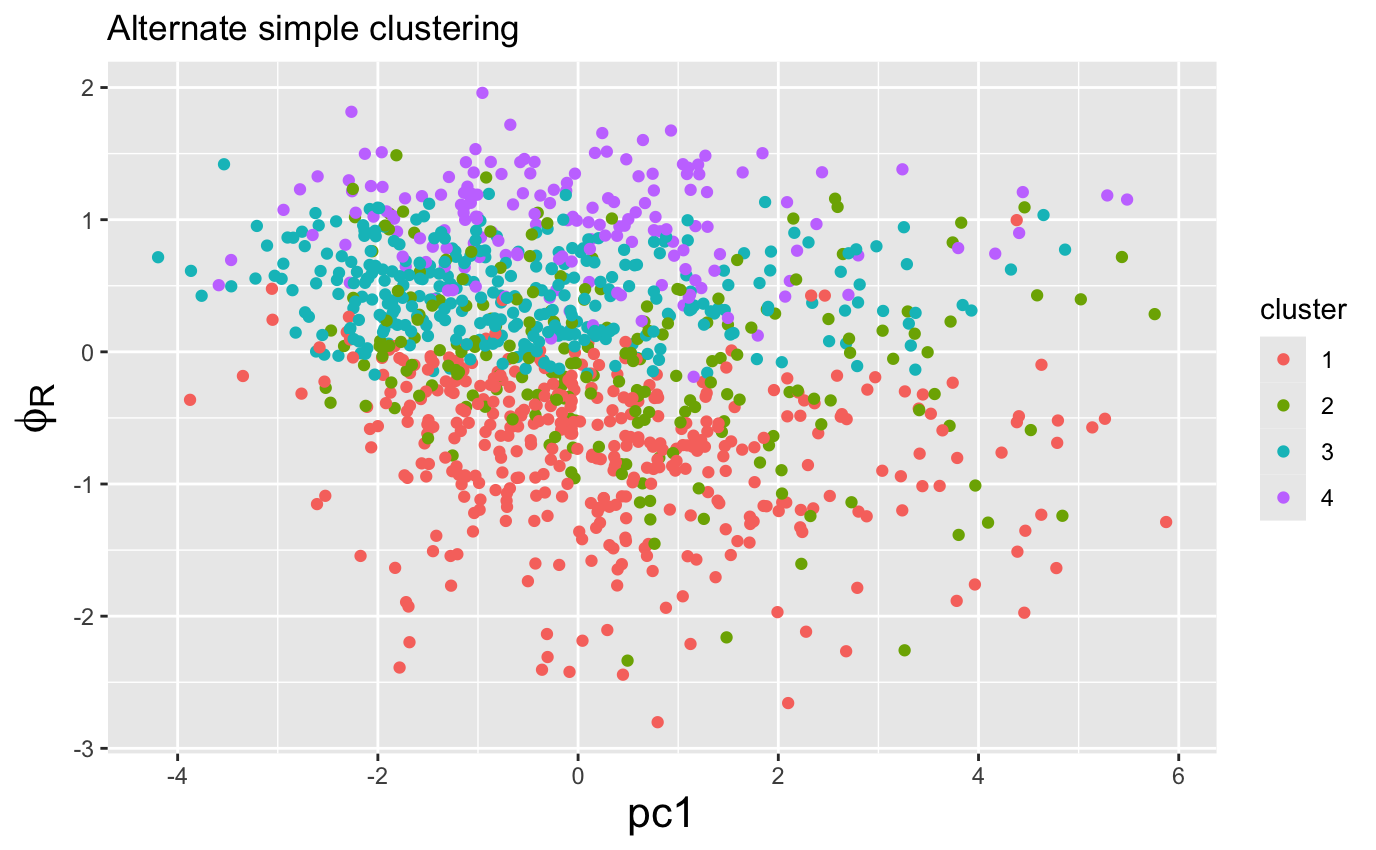}
\caption{\small{Cluster assignments computed using the K-means algorithm with REM AHI, non-REM AHI, time in REM, and time in non-REM as input variables, plotted on the same axes from Figure \ref{fig_cluster_2panel}.}}
\label{fig_alt_clustering}
\end{figure}

To get a sense of how these clusters compare with those estimated using our model random effects, Figure \ref{fig_alt_clustering} shows the cluster assignments visualized on the same axes of Figure \ref{fig_cluster_2panel}. Observe that clusters 1, 3, and 4 have a pattern similar to before along the $\phi_R$ axis, but now all four clusters are spread widely over the horizontal axis, indicating that these new clusters do not capture the variation in sleep dynamics that was captured by our model random effects.

Finally, we examine the differences in the BSRT scores among these alternate groups. Using the same linear regression setup as above, we find that $95\%$ credible intervals for all of the cluster coefficients include $0$, regardless of the choice of reference level. A full summary of the linear regression coefficients using cluster 4 as the reference is included in the Supplementary Information. Taken together, the above results suggest that in practice, simple summary statistics such as time in each sleep stage and stage-specific AHI may help give complementary insights to those generated by our model, but do not fully capture the same variability. More broadly, examining how other variables differ by the clusters generated by our model's random effects may help reveal other variables that could be key to better understanding sleep differences among OSA patients.

\section{Discussion}\label{section_discussion}

In this work, we investigated the heterogeneity in sleep stage dynamics and stage-specific rates of apnea and hypopnea events in OSA patients by analyzing data from the APPLES study, which generated several novel insights. We found that for the average patient, having an OSA event during a given epoch increases the probability of awakening or changing sleep stages in the following epoch, although both the baseline probabilities and the magnitude of this change vary substantially among patients. Also, for the average patient, the expected wait time between OSA events is substantially shorter during REM sleep than during non-REM, although again there is significant patient variation in rates. By modeling the covariance of our model's random effects, we revealed that during REM sleep, patient baseline rates of transitions to Awake or non-REM are positively correlated with one another, as well as being correlated with greater impact of OSA events on these transitions, while during non-REM sleep, baseline rates of transitions to REM or Awake are negatively correlated with the corresponding shifts due to events. Finally, we clustered patients into four interpretable groups based on their random effects. This revealed that other demographic and health variables, such as age, sex, and BMI, vary significantly by event rates and sleep dynamics. Turning to the cognitive performance measures BSRT and Pathfinder, we determined that relative to cluster 4, which had the highest average AHI, cluster 1 had better BSRT scores, but not better Pathfinder scores, after correcting for age and sex. This result was noteworthy because, while cluster 1 had the lowest rate of events during REM sleep, it did not have the lowest overall AHI, supporting the hypothesis that events in REM sleep may be particularly harmful to memory-dependent processes. In contrast to our clustering approach, a regression of BSRT on AHI directly yielded a $95\%$ credible interval that included zero, illustrating the limitations of an AHI-based analysis.

To accomplish the above, we developed a hierarchical Bayesian modeling approach for analyzing PSG data. We model OSA event rates and sleep stage dynamics jointly, including patient-specific effects of event occurrences on subsequent dynamics. We modeled the ten patient-level random effects controlling variation in these dynamics and event rates using a multivariate Gaussian distribution with factor-model covariance, thus allowing insights into shared characteristics in patient profiles. To summarize patients based on their estimated random effects, we showed that a Bayes-optimal clustering under K-means loss can be easily computed when clustering by random effects is desired. This clustering result, along with the associated uncertainty quantification we propose, may be of independent interest to readers, as it generalizes to any setting where one would like to cluster observations by random effects estimated using a Bayesian model. In simulations, we showed that our inferences are robust to model misspecification and that our clustering approach captures variation missed by a simpler clustering approach using only summary statistics of event rates and sleep architecture. Thus, while our model is a simplification of the true dynamics, ignoring nonstationarity of sleep dynamics and event rates, we argue it is sufficiently complex to generate useful inferences while maintaining both interpretability and computational feasibility.

There are a number of promising future directions. Both the inter-event time and dynamics models could be generalized. For example, the fixed and random effects for transitions between sleep stages could change as a function of time, reflecting that sleep dynamics differ between the beginning and end of the night. Similarly, the rates of event occurrences during REM and non-REM sleep could vary over time. In addition, the Exponential could be replaced with a more flexible alternative such as the Weibull to relax the memorylessness property. Adding such complexities to the model may more accurately approximate the true dynamics but also makes interpretation more challenging and may substantially complicate posterior sampling. Computational efficiency is already a substantial challenge given the immense size of the data, so developing a more efficient implementation of our current approach even without model elaborations would be worthwhile. One possibility is to model the random effect covariance as diagonal rather than using a factor model. In this case, each of the multinomial and Poisson models could be fit separately, taking only seconds to a few minutes. Although it does not allow for inferences about the covariance structure of $\theta_i$, this approach may be useful to practitioners looking to generate results more quickly, and still allows for the clustering approach to be applied in a second stage.

Finally, a separate direction is to extend our results for clustering based on posterior samples of model parameters. For example, loss functions other than K-means may be desirable in some situations; it would be interesting and useful to explore approaches for Bayes-optimal clustering in these more general settings. Finally, our approach could be applied to additional data sets to answer questions we did not address here. For example, while we considered only diagnostic PSG data for OSA patients, it would be interesting to explore how sleep stage dynamics parameters vary between OSA patients and healthy volunteers, as well as how they change for OSA patients in response to treatment, for example by a CPAP machine.

\section{Significance Statement}

\subsection*{The Problem}
Sleep apnea is diagnosed using the Apnea-Hypopnea Index (AHI), a count of breathing interruptions per hour. However, patients with identical AHI values often have vastly different symptoms, health risks, and treatment responses. Clinicians have long recognized this heterogeneity but lacked rigorous tools to quantify it beyond simple event counting.

\subsection*{Why This Matters}
This limitation prevents us from predicting which patients face highest cardiovascular risk, who will benefit from treatment, or why outcomes vary so dramatically. The field has called for "sleep apnea phenotyping" for over a decade, but without a principled statistical framework. Understanding heterogeneity is essential for personalized medicine and explaining inconsistent relationships between AHI and clinical outcomes.

\subsection*{Our Innovation}
We develop the first framework jointly modeling: (1) event rates during different sleep stages, (2) sleep stage transition dynamics, and (3) how much each patient's events disrupt their sleep architecture, we quantify patient-specific "disruption profiles" capturing not just event frequency but sleep fragmentation severity. We introduce a clustering approach to identify distinct phenotypes based on these profiles.

\subsection*{Advantages Over Existing Approaches}
Previous methods analyze event rates separately from sleep dynamics or use simple summary statistics. Our approach uniquely captures the bidirectional relationship: events affect sleep transitions, and this effect varies substantially by patients. Unlike AHI, which treats many mild events identically to fewer severe disruptive events, our model distinguishes these patterns with principled uncertainty quantification.

\subsection*{Key Findings}
We identified four distinct phenotypes: (1) low REM event rates with moderate disruption, (2) highly fragmented sleep despite moderate event rates, (3) stable sleep with low non-REM events, and (4) very high event rates with preserved sleep continuity. These phenotypes differ in BMI, demographics, and sleep architecture, differences missed by AHI. Exploratory analyses suggest phenotypes may relate to cognitive outcomes differently than event frequency alone.

\subsection*{Implications}
A prior phenotyping study \citep{zinchuk2018polysomnographic} identified sleep apnea subgroups using polysomnographic summary statistics and clinical features, demonstrating that phenotypes predict cardiovascular outcomes differently than AHI alone. Our framework advances this by modeling the dynamic relationship between OSA events and sleep architecture, providing patient-specific quantification of sleep disruption severity with continuous measures rather than categorical groupings.
This enables mechanistic research previously difficult: Does sleep fragmentation (independent of event frequency) drive cardiovascular risk? Do disruption patterns predict treatment response better than AHI? Our framework can test whether CPAP benefits work through reducing event frequency, reducing disruption per event, or stabilizing sleep architecture.

\subsection*{Path Forward}
While computationally intensive for immediate clinical use, this work establishes proof-of-concept for disruption-based phenotyping. Critical next steps include validating phenotypes in independent cohorts for cardiovascular outcomes and treatment response, assessing longitudinal stability of disruption patterns, and developing simplified scoring systems for clinical implementation. If validated, this approach could transform sleep apnea assessment from event counting to personalized disruption profiling.

\section{Acknowledgments}

The authors gratefully acknowledge funding from the National Institutes of Health (NIH) grant R01ES035625, the Office of Naval Research (ONR) grant N00014-21-1-2510, and Merck \& Co., Inc., through its support for the Merck BARDS Academic Collaboration. The authors also acknowledge the Duke Compute Cluster for computational time and Surya Tokdar for helpful discussions.

\bibliographystyle{chicago}
\bibliography{ref}

\end{document}


\maketitle

\section{Proof of Proposition 3.1}

Observe that, letting all expectations be over the posterior distribution $\pi_{post}$, we have
\begin{align*}
    \left(\hat{c}, \hat{b}\right) &= \argmin_{c: |C|=K, \,\, b \in \mathbb{R}^{K \times d}} \sum_{k=1}^K \sum_{i \in C_k} \mathbb{E} \left[||\theta_i - b_k||^2 \right] \\
    &= \argmin_{c: |C|=K, \,\, b \in \mathbb{R}^{K \times d}} \sum_{k=1}^K \sum_{i \in C_k} \mathbb{E} \left[||(\theta_i - \mathbb{E}[\theta_i]) + (\mathbb{E}[\theta_i] - b_k)||^2 \right] \\
    &= \argmin_{c: |C|=K, \,\, b \in \mathbb{R}^{K \times d}} \sum_{k=1}^K \sum_{i \in C_k} \mathbb{E}[||\theta_i - \mathbb{E}[\theta_i]||^2] + \mathbb{E}[(\theta_i - \mathbb{E}[\theta_i])^T (\mathbb{E}[\theta_i] - b_k)] + \mathbb{E}[||\mathbb{E}[\theta_i] - b_k||^2] \\
    &= \argmin_{c: |C|=K, \,\, b \in \mathbb{R}^{K \times d}} \sum_{k=1}^K \sum_{i \in C_k} ||\mathbb{E}[\theta_i] - b_k||^2
\end{align*}
where we have used that in the second-to-last line, the first term does not depend on $(c,b)$, the second term is equal to zero, and the outer expectation in the third term can be dropped. $\square$

\section{Summary of important abbreviations and model parameters}

\newpage
\begin{table}[h!]
\centering
\begin{tabular}{|l|l|}
\hline
\textbf{Term}      & \textbf{Meaning}                                                                                                                                                                                                                                                                                                                              \\ \hline
OSA                & \begin{tabular}[c]{@{}l@{}}Obstructive Sleep Apnea: A breathing disorder\\ characterized by periodic narrowing or obstruction\\ of the airway during sleep\end{tabular}                                                                                                                                                                       \\ \hline
PSG                & \begin{tabular}[c]{@{}l@{}}Polysomnogram: An overnight sleep study that\\ collects a range of physiological information\end{tabular}                                                                                                                                                                                                          \\ \hline
AHI                & \begin{tabular}[c]{@{}l@{}}Apnea-Hypopnea Index: Calculated as \\ $\frac{\text{\#apnea events + \#hypopnea events}}{\text{\# hours sleep}}$\end{tabular}                                                                                                                                                                                         \\ \hline
REM                & Rapid Eye Movement sleep                                                                                                                                                                                                                                                                                                                      \\ \hline
non-REM            & \begin{tabular}[c]{@{}l@{}}Sleep other than REM sleep. (N1, N2, and N3\\ refer to light, medium, and deep non-REM sleep.)\end{tabular}                                                                                                                                                                                                        \\ \hline
APPLES             & \begin{tabular}[c]{@{}l@{}}The Apnea Positive Pressure Long-Term Efficacy\\ Study\end{tabular}                                                                                                                                                                                                                                                \\ \hline
BSRT               & \begin{tabular}[c]{@{}l@{}}The Buschke Verbal Selective Reminding Test: \\ Patients are asked to recall a set of words after \\ varying intervals of time. The score is the total \\ number of words recalled correctly.\end{tabular}                                                                                                         \\ \hline
Pathfinder         & \begin{tabular}[c]{@{}l@{}}The Pathfinder Number Test: Patients select \\ consecutive numbers in boxes on a computer \\ screen. The score is the number of seconds taken \\ to complete the task.\end{tabular}                                                                                                                                \\ \hline
$\mu_{s_o s_n}$    & \begin{tabular}[c]{@{}l@{}}Model parameter controlling the baseline rate of \\ transitions from sleep stage $s_o \in \{\text{REM, non-REM}\}$ \\ to stage $s_n \in \{\text{Awake, REM, non-REM}\}$\end{tabular}                                                                                                                               \\ \hline
$\gamma_{is_os_n}$ & \begin{tabular}[c]{@{}l@{}}Mean zero random effect controlling how patient\\ $i$'s rate of transitions from stage $s_o$ to $s_n$\\ varies away from $\mu_{s_os_n}$\end{tabular}                                                                                                                                                               \\ \hline
$\tau_{s_o s_n}$   & \begin{tabular}[c]{@{}l@{}}Model parameter controlling how the rate of transitions from stage\\ $s_o$ to $s_n$ shifts when a patient has an OSA event.\end{tabular}                                                                                                                                                                           \\ \hline
$\alpha_{is_os_n}$ & \begin{tabular}[c]{@{}l@{}}Mean zero random effect controlling how patient $i$'s shift due\\ to an OSA event varies away from $\tau_{s_os_n}$\end{tabular}                                                                                                                                                                                    \\ \hline
$\lambda_s$        & \begin{tabular}[c]{@{}l@{}}Model parameter controlling the expected time between OSA events\\ during sleep stage $s$. (The expected inter-event time for the\\ average patient is $1 / \lambda_s$.)\end{tabular}                                                                                                                              \\ \hline
$\phi_{is}$        & \begin{tabular}[c]{@{}l@{}}Mean zero random effect controlling patient variation in rates of\\ events in stage $s$. (The expected inter-event time for patient $i$\\ is $1 / (\lambda_s \text{exp}(\phi_{is}))$\end{tabular}                                                                                                                  \\ \hline
$\theta_i$         & \begin{tabular}[c]{@{}l@{}}The 10-dimensional vector of all random effects for patient $i$:\\ $\theta_i = (\gamma_{iRA}, \gamma_{iRN}, \gamma_{iNA}, \gamma_{iNR}, \alpha_{iRA}, \alpha_{iRN}, \alpha_{iNA}, \alpha_{iNR}, \phi_{iR}, \phi_{iN})^T$\\ where $R$ indicates REM, $N$ indicates non-REM, and\\ $A$ indicates Awake.\end{tabular} \\ \hline
\end{tabular}
\caption{\small{Summary of important abbreviations and model parameters.}}
\end{table}

\newpage

\section{Alternate clustering methods}

\subsection{Point estimation using the formulation in equation (2)}

For comparison to the clustering point estimate computed in the main paper based on equation (1), we compute an estimate based on equation (2), where we optimize only over cluster assignments, not simultaneously estimating the cluster centers. Recall the form of equation (2):
\begin{equation}\label{opt_simple}
\tag{S1}
    \hat{c}' = \argmin_{c: |C|=K} \sum_{k=1}^K \sum_{i \in C_k} \mathbb{E}_{\pi_{post}} \left[||\theta_i - \overline{\theta}_k||^2 \right],
\end{equation}
where $\overline{\theta}_k$ is the mean of the observations in cluster $k$. Observe that to compute $\hat{c}'$ in the above given $m$ posterior samples of each $\theta_i$ vector, we can again apply standard K-means code, but using $\Tilde{\theta}_i \in \mathbb{R}^{pm}$ for $i=1,...,n$ as input data, where we define $\Tilde{\theta}_i$ as all $m$ posterior samples of $\theta_i \in \mathbb{R}^p$ concatentated together into a single vector. To see why, observe that using posterior means to estimate the expectations in \eqref{opt_simple}, the loss function becomes
\begin{align*}
    \sum_{k=1}^K \sum_{i \in C_k} \sum_{j=1}^m ||\theta_{ij} - \overline{\theta}_{kj}||^2 &= \sum_{k=1}^K \sum_{i \in C_k} \sum_{j=1}^m \sum_{l=1}^p (\theta_{ijl} - \overline{\theta}_{kjl})^2 \\
    &= \sum_{k=1}^K \sum_{i \in C_k} ||\Tilde{\theta}_i - \overline{\Tilde{\theta}}_k||^2,
\end{align*}
i.e., the standard K-means loss function applied to vectors $\Tilde{\theta}_1,...,\Tilde{\theta}_n$. Implementing this in R \citep{R_lang}, we get the estimated cluster assignments shown in Figure \ref{fig_alt_clustering_point_estimate} below, visualized on the same axes as in Figure 4 of the paper. Ignoring label switching, the clusters appear very similar to the ones we computed using equation (1). The adjusted Rand index between the two clusterings is 0.998, indicating extremely strong agreement. Thus, for this setting, the two estimates are almost interchangeable. We suggest a more full comparison of the two approaches as an interesting direction for future work.

\begin{figure}[H]
\centering
\includegraphics[width=0.75\textwidth]{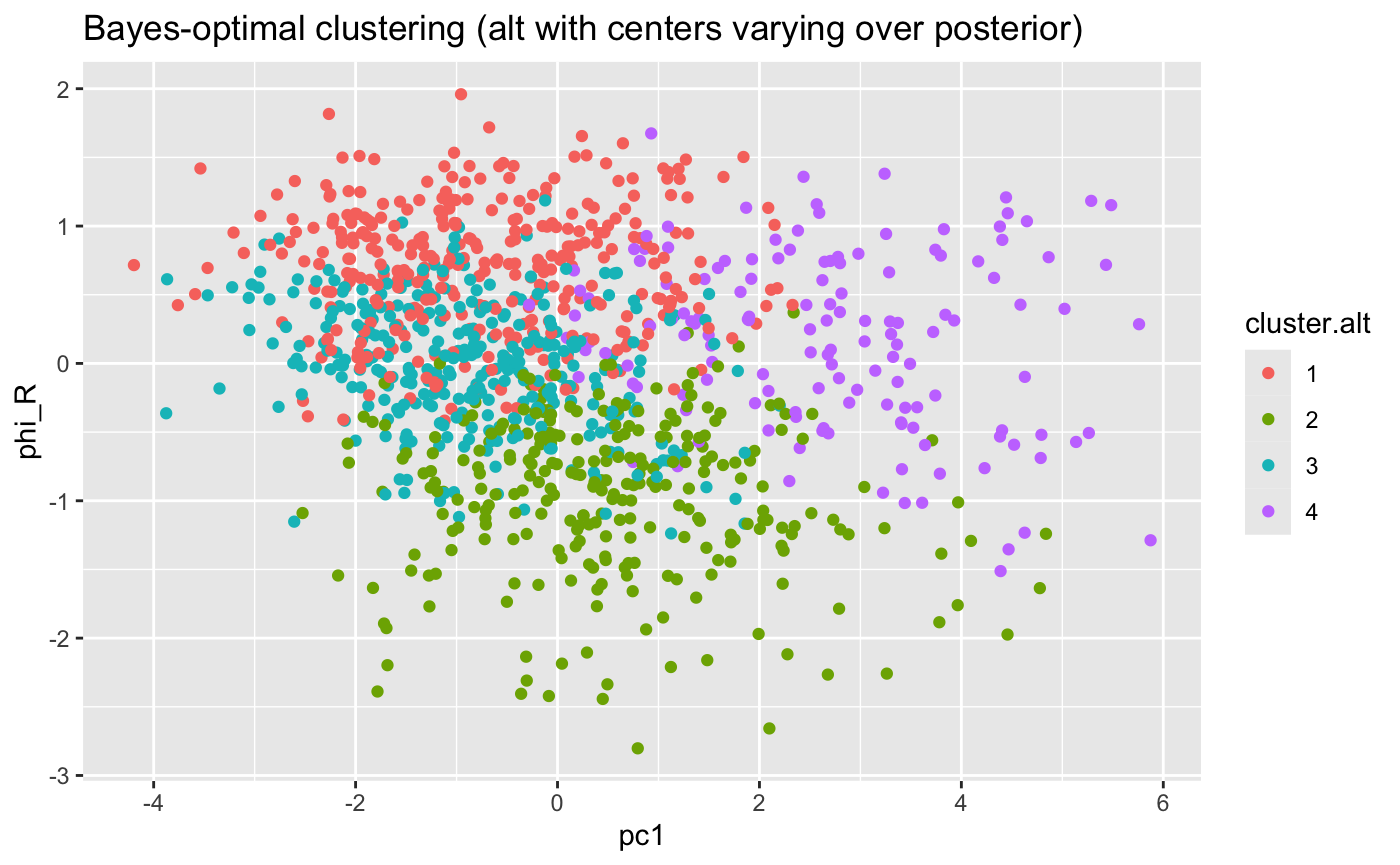}
\caption{\small{Bayes-optimal k-means point estimate of cluster assignments based on random effect vectors $\theta_i$ using the formulation in \eqref{opt_simple}. pc1 refers to the first principal component vector computed with the posterior means of the dynamics model random effects.}}
\label{fig_alt_clustering_point_estimate}
\end{figure}

\subsection{Alternate approaches for clustering uncertainty quantification}

\subsubsection{Applying k-means to each posterior sample}

One conceptually simple approach to characterizing uncertainty in the optimal clusters and centers is to simply run the k-means algorithm separately on each posterior sample, generating $m$ posterior samples of the optimal clustering.
\begin{equation}\label{kmeans_samples}
\tag{S2}
    (\Tilde{c}_j, \Tilde{b}_j) = \argmin_{c: |C|=K, \,\, b \in \mathbb{R}^{K \times d}} \sum_{k=1}^K \sum_{i \in C_k}  ||\theta_{ij} - b_{kj}||^2, \,\,\,\,\,\,\,\,\,\,\,\,\,\,\,\,\, j=1,...,m. 
\end{equation}
While this is in some sense a natural choice, it presents two practical issues. First, relating summaries of these clusterings directly to our point estimate is nontrivial due to label-switching, so we would instead need to rely on related summaries like pairwise co-clustering probabilities. Second, and more fundamentally, the relationship between the distribution of the $\argmin$s in \eqref{kmeans_samples} and the $\argmin$ of the expected loss in equation (1) is also not straightforward; e.g., our point estimate using equation (1) may be assigned a posterior probability of $0$ in \eqref{kmeans_samples}, while cluster assignments sampled in \eqref{kmeans_samples} may be extremely suboptimal in equation (1), making it difficult to interpret summaries this method produces.

Despite these difficulties, we implemented this approach, and we summarize the results below in Figure \ref{fig_alt_uq_separate_kmeans}. Pairwise co-clustering probabilities for all patients are visualized in the left panel of the figure, with dark red representing a probability of $1$, and light yellow representing $0$. The observations are ordered so that the observed block-diagonal pattern corresponds to observations whose point estimates were cluster 1, 2, 3, and 4. Clearly, there is some general agreement between the co-clustering probabilities and the point estimate; however, there are also a number of patients whose co-clustering probabilities with others in their cluster point estimate are quite low. Since there is no reliable notion of cluster locations across samples now, to summarize the probabilities of assignment to ``Cluster 1,'' we computed co-clustering probabilities with the patient whose $\theta_i$ posterior mean was the closest to our cluster 1 center point estimate. Doing so yields a similar distribution of probabilities as in Figure 4 in the paper; however, it is misleading to give the values the same interpretation, as they are computed in fundamentally different ways. 

\begin{figure}[H]
\centering
\includegraphics[width=\textwidth]{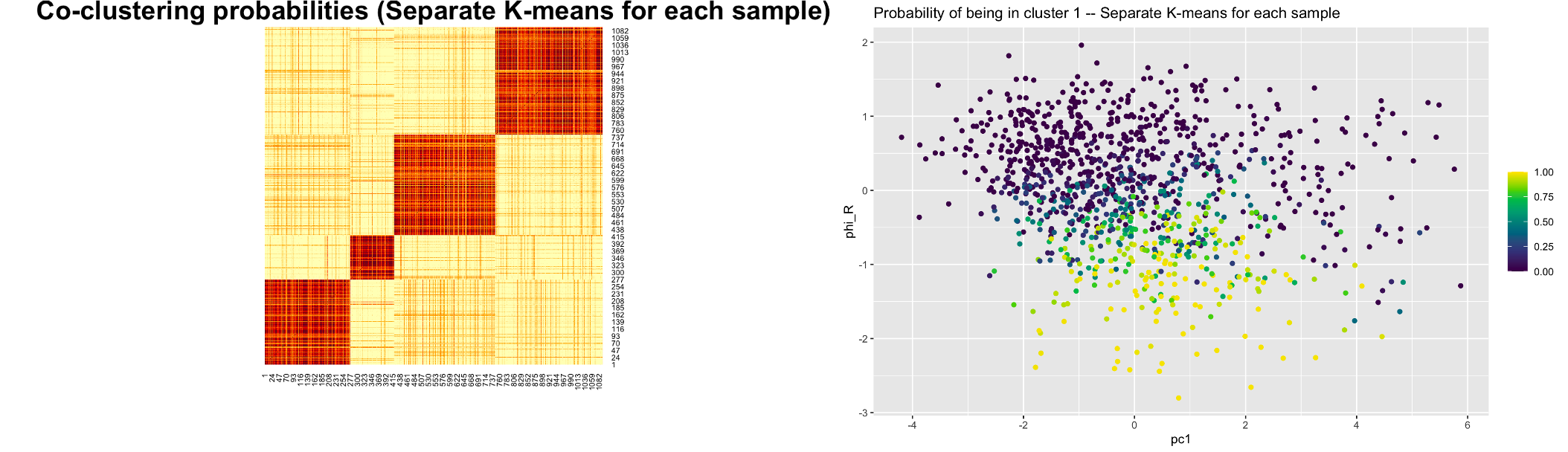}
\caption{\small{Co-clustering probabilities (left) and probabilities of being co-clustered with the individual closest to the cluster 1 center point estimate (right) for the distribution of clusterings learned by applying K-means to each of the 10,000 posterior samples of $\theta_1,...,\theta_n$.}}
\label{fig_alt_uq_separate_kmeans}
\end{figure}

Finally, to compare this learned distribution of clusterings to our point estimate directly, we use the adjusted Rand index between each of the 10,000 cluster estimates and our point estimate. This results in a mean ARI of $0.58$, standard deviation $0.05$, and a range of $0.28$ to $0.71$. This suggests meaningful disagreement on average, and none of the individual sample clusterings are equivalent to the point estimate. 

\subsubsection{Generalized Bayesian inference based on \cite{rigon2023generalized}}

A second approach for loss-based clustering uncertainty quantification is the generalized Bayes approach proposed by \cite{rigon2023generalized}. If we ignore patient-level uncertainty and treat each $\hat{\theta}_i$ as observed, we can sample from a Gibbs posterior representing rational posterior beliefs about the optimal clustering in terms of expected k-means loss over the unknown population distribution of $\theta_i$, again yielding a distribution of clusterings. In addition to the difficulties in summarizing such a distribution mentioned above, this Gibbs posterior critically depends on a tuning parameter $\lambda$; at extreme choices of $\lambda$, the clustering distribution can collapse to a point mass at a single paritition, or can be flat over all possible partitions. While \cite{rigon2023generalized} propose an approach for choosing a reasonable $\lambda$ for k-means, interpretability of the resulting probabilities is again a concern.

We also implement this approach for comparison, using the Gibbs sampler described in \cite{rigon2023generalized} with 10,000 samples drawn. We show the same summaries as in the previous section below in Figure \ref{fig_alt_uq_gibbs}. In general, agreement with our point estimate appears stronger here than in Figure \ref{fig_alt_uq_separate_kmeans}. 

\begin{figure}[H]
\centering
\includegraphics[width=\textwidth]{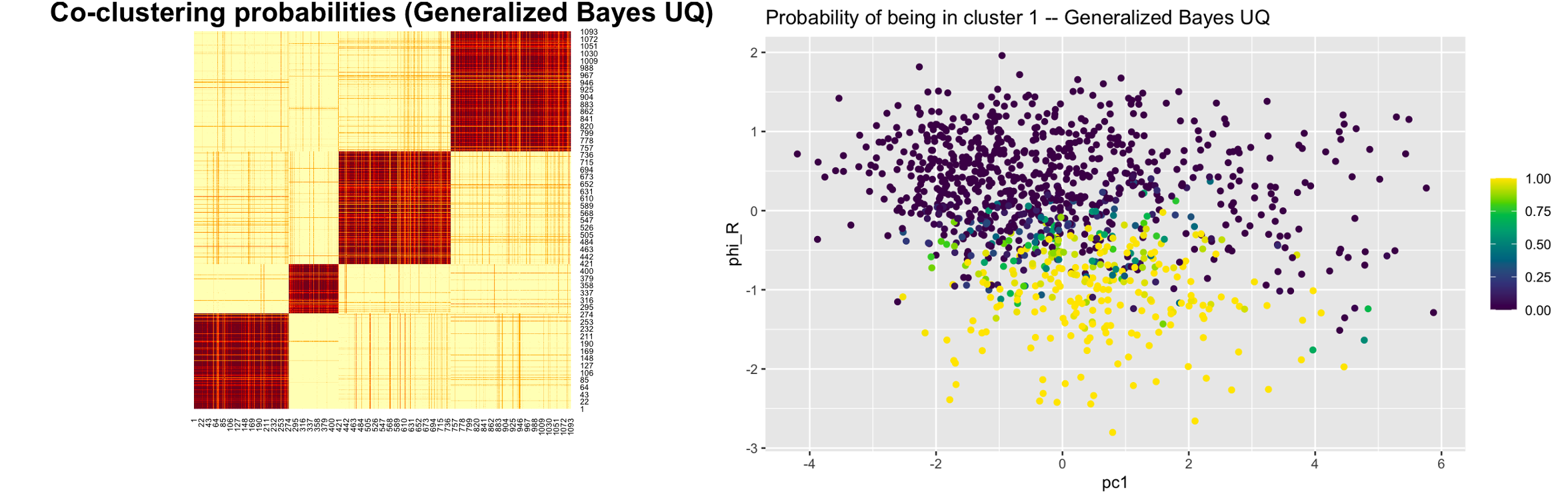}
\caption{\small{Co-clustering probabilities (left) and probabilities of being co-clustered with the individual closest to the cluster 1 center point estimate (right) for the Gibbs posterior of clusterings learned by applying the method of \cite{rigon2023generalized}.}}
\label{fig_alt_uq_gibbs}
\end{figure}

Computing ARI values between each sample and point estimate, we get a mean of $0.76$, standard deviation $0.02$, and a range of $0.69$ to $0.83$. Thus, again we get that our point estimate has an estimated probability of zero. 

\section{Additional posterior predictive checks}

\begin{figure}[H]
\centering
\includegraphics[width=\textwidth]{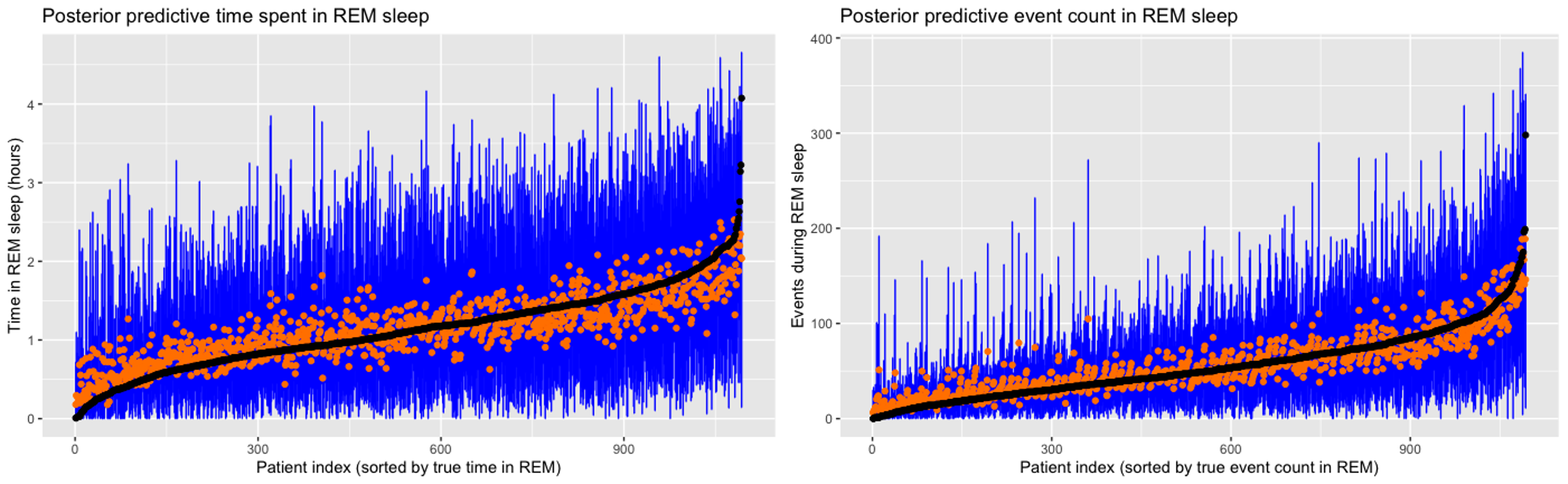}
\caption{\small{Posterior predictive distributions of time spent in REM sleep (left) and the number of events occurring during REM sleep (right). Black dots are observed values; orange dots are posterior predictive means; blue lines are $95\%$ posterior predictive intervals.}}
\label{fig_post_pred_REM}
\end{figure}

\begin{figure}[H]
\centering
\includegraphics[width=\textwidth]{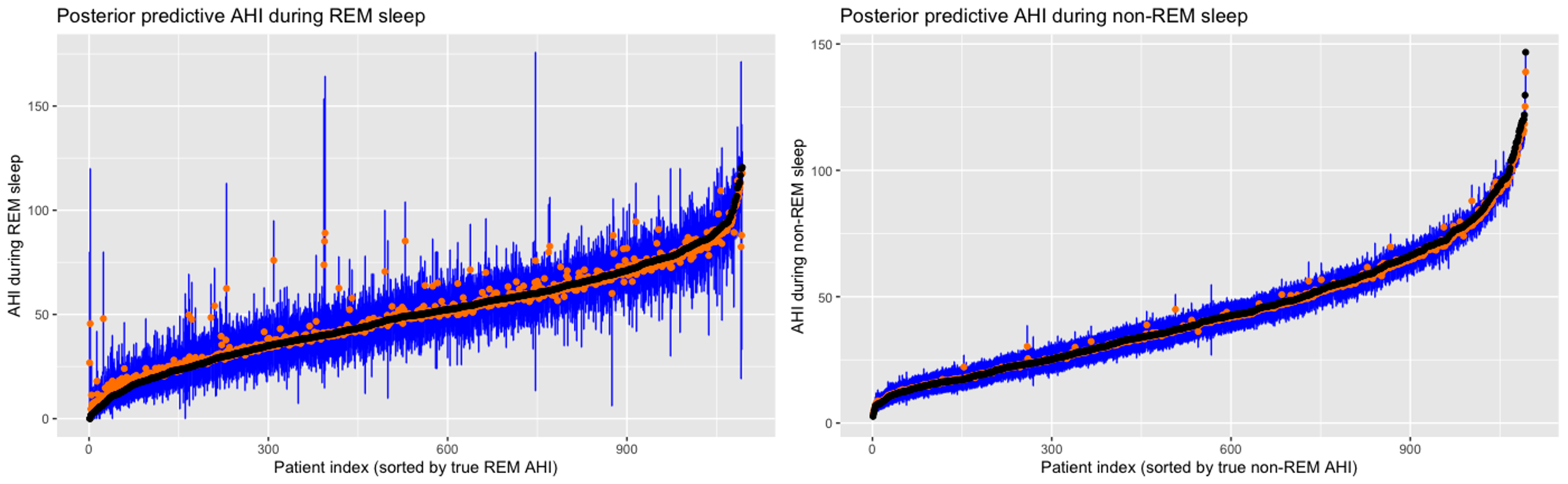}
\caption{\small{Posterior predictive distributions of AHI during REM sleep (left) and AHI during non-REM sleep (right). Black dots are observed values; orange dots are posterior predictive means; blue lines are $95\%$ posterior predictive intervals.}}
\label{fig_post_pred_AHI}
\end{figure}

\section{EDA figures}

\begin{figure}[H]
\centering
\includegraphics[width=0.75\textwidth]{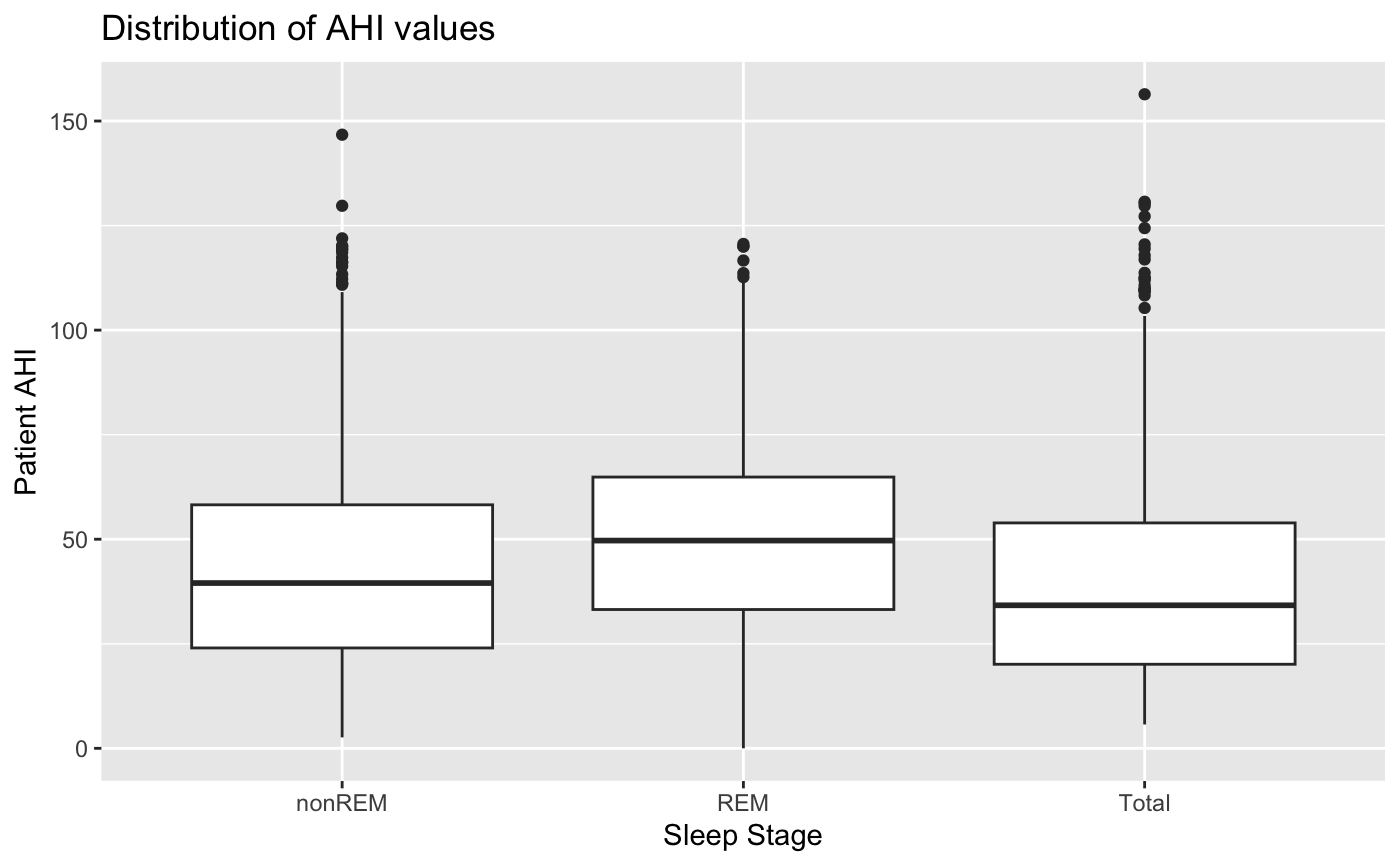}
\caption{\small{Distribution of REM, non-REM, and overall AHI values for patients in the APPLES study.}}
\label{fig_AHI_boxplot}
\end{figure}

\begin{figure}[H]
\centering
\includegraphics[width=0.75\textwidth]{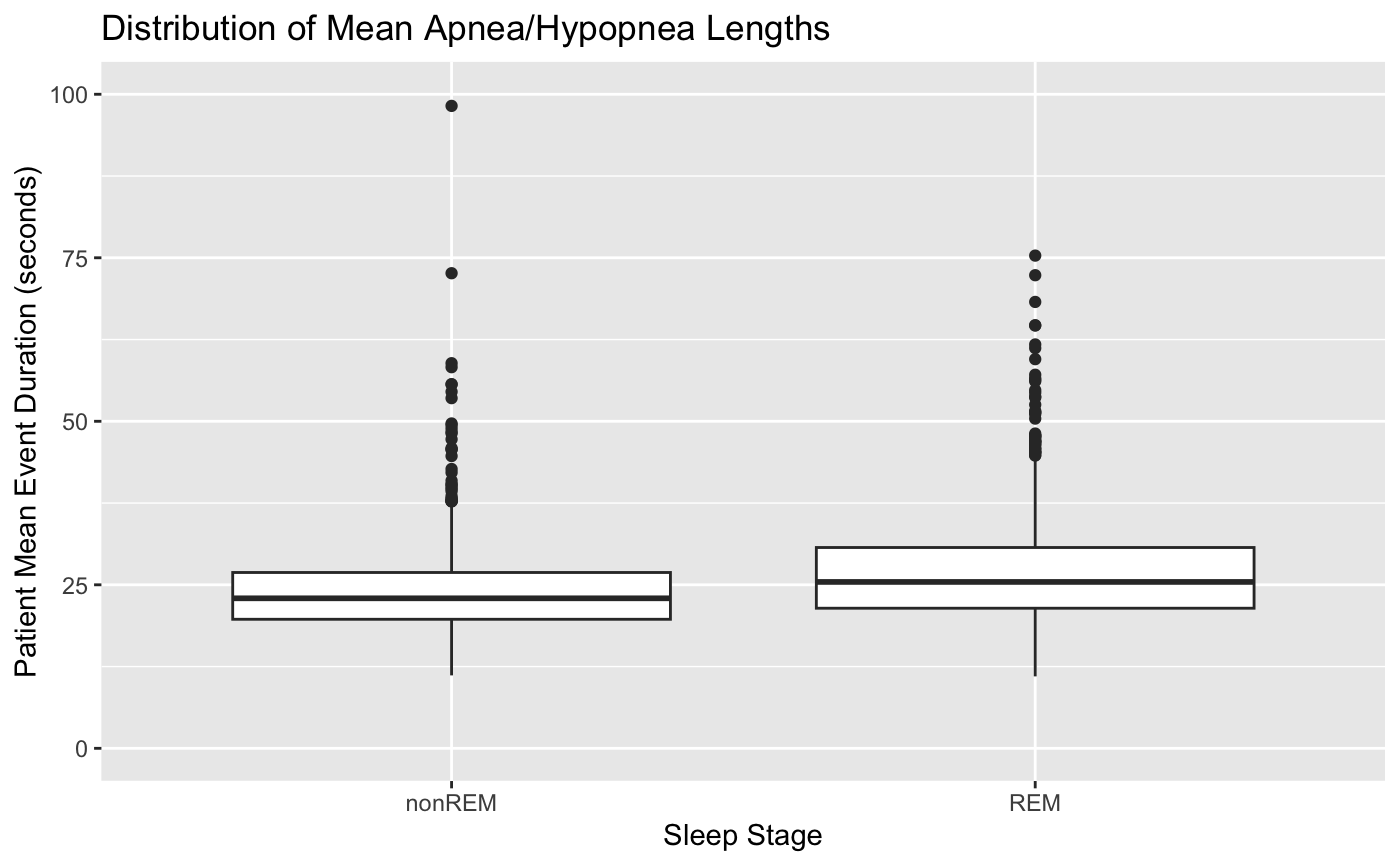}
\caption{\small{Distribution of average hypopnea/apnea event durations for patients in the APPLES study. Four patients are omitted for REM sleep: two with zero events during REM, and two with outlier mean event lengths of 219 and 188 seconds.}}
\label{fig_event_lengths_boxplot}
\end{figure}

\begin{figure}[H]
\centering
\includegraphics[width=0.8\textwidth]{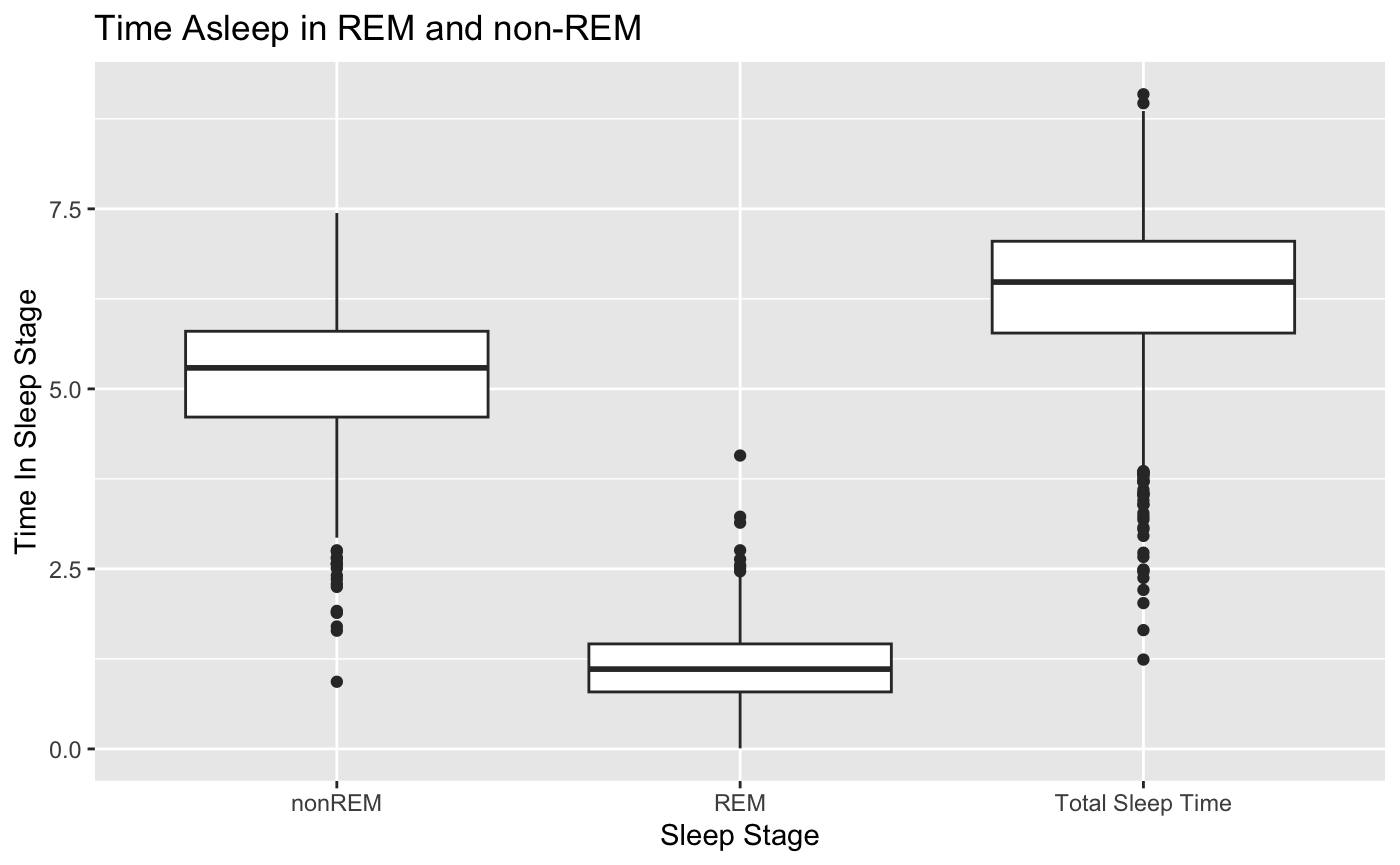}
\caption{\small{Distributions of time in non-REM sleep, time in REM sleep, and total time asleep in hours for patients in the APPLES study.}}
\label{fig_time_asleep_boxplot}
\end{figure}

\section{Aligned factor loadings}

\begin{figure}[H]
\centering
\includegraphics[width=0.8\textwidth]{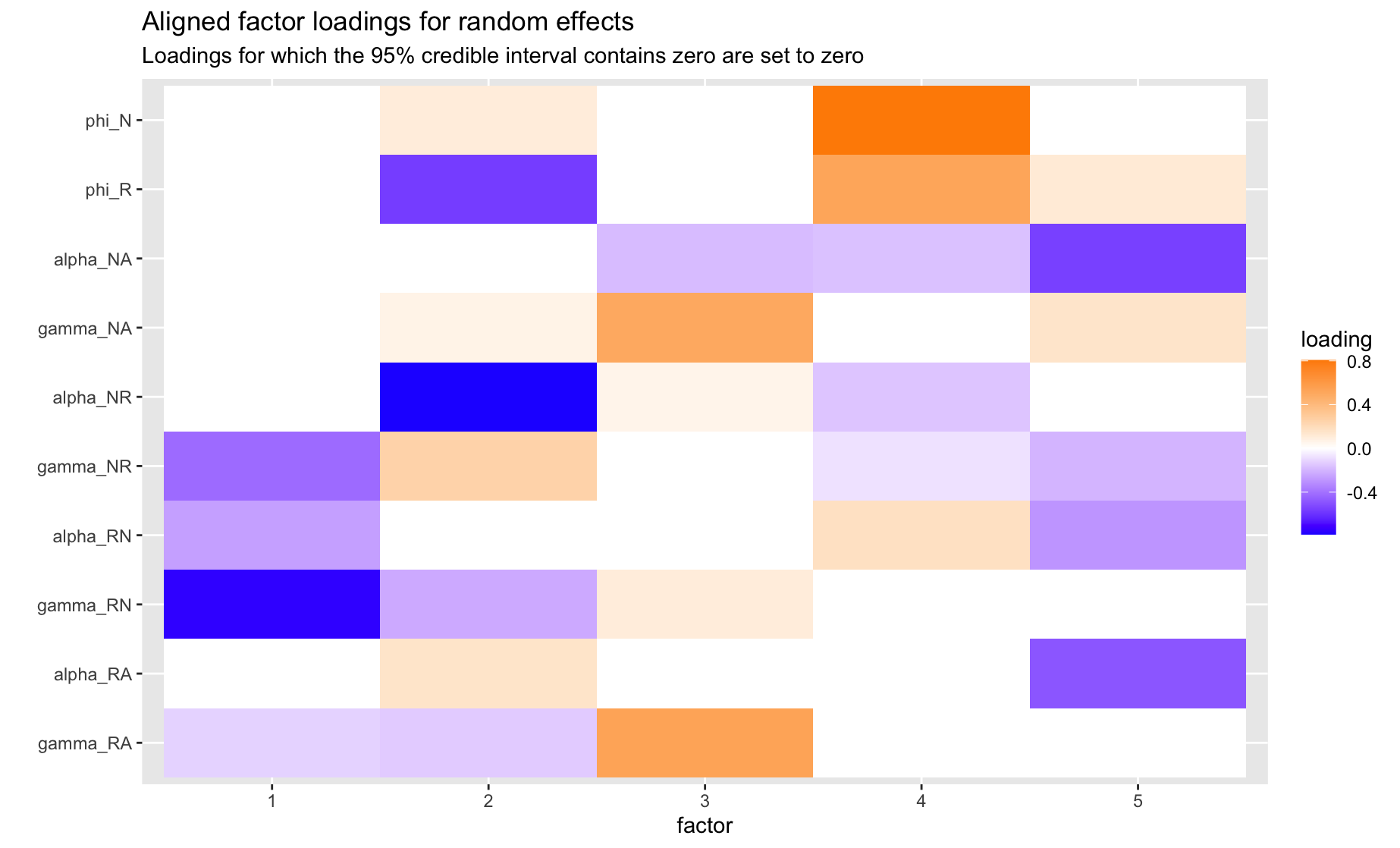}
\caption{\small{Posterior mean factor loadings after alignment with the procedure of \cite{poworoznek2021efficiently}. Loadings for which the $95\%$ credible interval contains zero have been set to zero.}}
\label{fig_aligned_factors}
\end{figure}

\section{BSRT regression on simple alternate clusters}


\begin{table}[H]
\centering
\begin{tabular}{llll}
\hline
                              & Coefficient & Posterior Mean & 95\% Posterior Credible Interval \\ \hline
\multirow{6}{*}{Regression 3} & Intercept   & \textbf{62.56} & \textbf{(59.96, 65.19)}          \\
                              & Cluster1    & 1.45           & (-0.11, 3.01)                    \\
                              & Cluster2    & 1.45           & (-0.36, 3.18)                    \\
                              & Cluster3    & 0.28           & (-1.32, 1.86)                    \\
                              & Age         & \textbf{-0.21} & \textbf{(-0.25, -0.17)}          \\
                              & SexM        & \textbf{-4.09} & \textbf{(-5.20, -2.99)}          \\ \cline{2-4} 
\end{tabular}
\caption{\small{Results for a linear regression of BSRT score on alternate cluster labels with reference level 4, after accounting for age and sex. Clusters were computed using REM AHI, non-REM AHI, time in REM sleep, and time in non-REM sleep, and are summarized in the paper.}}
\label{table_regression_alt}
\end{table}

\bibliographystyle{chicago}
\bibliography{ref}